\documentclass[fleqn,usenatbib]{mnras}

\usepackage{newtxtext,newtxmath}
\usepackage{longtable}

\usepackage[T1]{fontenc}

\DeclareRobustCommand{\VAN}[3]{#2}
\let\VANthebibliography\thebibliography
\def\thebibliography{\DeclareRobustCommand{\VAN}[3]{##3}\VANthebibliography}

\usepackage{graphicx}	
\usepackage{amsmath}	

\def\rhouv{$\rho_{\mathrm{UV}}$}
\newcommand{\muv}{$M_{\mathrm {UV}}$}

\def\msol{M$_{\odot}$}

\newcommand{\jwst}{{\em JWST}}
\newcommand{\hst}{{\em HST}}

\def\rhouv{$\rho_{\mathrm{UV}}$}
\def\rhosfr{$\rho_{\rm SFR}$}
\newcommand{\ra}[3]{#1$^{\mathrm{h}}$#2$^{\mathrm{m}}$#3$^{\mathrm{s}}$}
\newcommand{\decl}[3]
{#1$^{\circ}$#2$'$#3$''$}

\title[A Glimpse of the ultra-faint galaxies at Cosmic Dawn]{The first GLIMPSE of the faint galaxy population at Cosmic Dawn with JWST: The evolution of the ultraviolet luminosity function across $z\sim9-15$}

\author[Chemerynska et al.]{Iryna Chemerynska,$^{1}$\thanks{E-mail: iryna.chemerynska@iap.fr}
Hakim Atek,$^{1}$
Lukas J. Furtak,$^{2}$
John Chisholm,$^{3,4}$
Ryan Endsley,$^{3}$
Vasily Kokorev,$^{3}$\newauthor
Joki Rosdahl,$^{5}$
Jeremy Blaizot,$^{5}$
Angela Adamo,$^{6}$
Rychard Bouwens,$^{7}$
Seiji Fujimoto,$^{3}$
Damien Korber,$^{7}$\newauthor
Charlotte Mason,$^{9}$
Kristen B.~W.\ McQuinn,$^{10,11}$
Julian B.~Mu\~noz,$^{3}$
Priyamvada Natarajan,$^{12,13}$\newauthor
Erica Nelson,$^{14}$
Pascal A. Oesch,$^{8,9}$
Richard Pan,$^{15}$
Johan Richard,$^{16}$
Alberto Saldana-Lopez,$^{5}$\newauthor
Marta Volonteri,$^{1}$
Adi Zitrin,$^{2}$
Danielle A. Berg,$^{3,4}$
Adélaïde Claeyssens,$^{5}$
Miroslava Dessauges-Zavadsky,$^{7}$\newauthor
Michelle Jecmen,$^{3,4}$
Ivo Labb\'e,$^{17}$
Rohan Naidu,$^{18}$
Maxime Trebitsch$^{19}$
\\
$^{1}$Institut d'Astrophysique de Paris, CNRS, Sorbonne Universit\'e, 98bis Boulevard Arago, 75014, Paris, France\\
$^{2}$Department of Physics, Ben-Gurion University of the Negev, P.O. Box 653, Be'er-Sheva 84105, Israel\\
$^{3}$Department of Astronomy, The University of Texas at Austin, Austin, TX 78712, USA\\
$^{4}$Cosmic Frontier Center, The University of Texas at Austin, Austin, TX 78712, USA \\
$^{5}$Universite Claude Bernard Lyon 1, CRAL UMR5574, ENS de Lyon, CNRS, Villeurbanne, F-69622, France\\
$^{6}$Department of Astronomy, The Oskar Klein Centre, Stockholm University, AlbaNova, SE-10691 Stockholm, Sweden\\
$^{7}$Leiden Observatory, Leiden University, NL-2300 RA Leiden, Netherlands\\
$^{8}$D\'epartement d'Astronomie, Universit\'e de Gen\`eve, Chemin Pegasi 51, 1290 Versoix, Switzerland\\
$^{9}$Cosmic Dawn Center (DAWN), Niels Bohr Institute, University of Copenhagen, Jagtvej 128, K{\o}benhavn N, DK-2200, Denmark\\
$^{10}$Space Telescope Science Institute, 3700 San Martin Dr., Baltimore, MD 21218, USA\\
$^{11}$Department of Physics \& Astronomy, Rutgers, The State University of New Jersey, Piscataway, NJ 08854, USA\\
$^{12}$Department of Astronomy, Yale University, 52 Hillhouse Ave, New Haven, CT 06511, USA\\
$^{13}$Black Hole Initiative, Harvard University, 20 Garden Street, Cambridge, MA 02138, USA\\
$^{14}$Department for Astrophysical and Planetary Science, University of Colorado, Boulder, CO 80309, USA\\
$^{15}$Department of Physics \& Astronomy, Tufts University, MA 02155, USA\\
$^{16}$Univ Lyon, Univ Lyon1, Ens de Lyon, CNRS, CRAL UMR5574, F-69230, Saint-Genis-Laval,France\\
$^{17}$Centre for Astrophysics and Supercomputing, Swinburne University of Technology, Melbourne, VIC 3122, Australia\\
$^{18}$MIT Kavli Institute for Astrophysics and Space Research, 70 Vassar Street, Cambridge, MA 02139, USA\\
$^{19}$LERMA, Sorbonne Université, Observatoire de Paris, PSL Research University, CNRS, 75014 Paris, France\\
}

\date{Accepted XXX. Received YYY; in original form ZZZ}

\pubyear{YYYY}

\begin{document}
\label{firstpage}
\pagerange{\pageref{firstpage}--\pageref{lastpage}}
\maketitle

\begin{abstract} 
Using ultra-deep \jwst /NIRCam imaging from the GLIMPSE Survey, enhanced by gravitational lensing of the Abell S1063 cluster, we investigate the faintest galaxies ever observed in the redshift range $z\sim9$ to $z\sim15$. We identify 105 galaxy candidates within this range, spanning absolute ultraviolet (UV) magnitudes from \muv$\sim-18$ to \muv$\sim-13$ mag, about three magnitudes fainter, on average, than prior \jwst\ studies. We place strong constraints on the ultra-faint end of the UV luminosity function (UVLF), finding minimal evolution in the faint-end slope, which varies from $\alpha=-2.01\pm 0.20$ at $z=9$ 
to $\alpha=-2.10\pm0.19$ at $z=13$ 
. This behaviour contrasts with the rapid evolution of the faint-end slope observed from $z \sim 0$ to $z \sim 9$. By integrating the UVLF down to \muv$=-16$, we derive the cosmic star formation rate (SFR) density, \rhosfr, from $z=9$ to $z=$ 13, revealing a best-fit redshift evolution that follows $\propto (1+z)^{-2.94^{+0.06}_{-0.10}}$. This slope is significantly shallower than predictions from most theoretical models. Extending the integration limit to \muv$=-13$, we find that galaxies fainter than \muv$=-16$ contribute more than 50\% of the total cosmic SFR density at $z\sim12$. The observed excess in the cosmic SFR density at these high redshifts may suggest an enhancement in the star formation efficiency during the earliest phases of galaxy formation. Alternatively, this could result from other physical mechanisms, such as bursty star formation histories; minimal dust attenuation; or an evolving initial mass function. However, existing models that incorporate these scenarios fail to fully reproduce the observed redshift evolution of \rhosfr. Finally, we acknowledge the potential impact of low-redshift contamination and cosmic variance, as the small survey volume may not represent the broader galaxy population. Similar observations in different fields and spectroscopic confirmation are required to validate these findings. 
\end{abstract}



\section{Introduction}\

One of the central challenges in modern astrophysics is to understand how the earliest galaxies and cosmic structures formed from primordial gas just a few hundred million years after the Big Bang \citep{dayal18}. The initial fluctuations in the matter power spectrum, determined by the cosmological densities of baryonic and non-baryonic matter, seed the formation of dark matter halos. These halos grow hierarchically by accreting material from the intergalactic medium, ultimately leading to the formation of stars \citep{press74, springel05}. Beyond the initial efficiency of star formation, stellar feedback plays a key role in regulating ongoing star formation by reducing the gas available for accretion and subsequent star formation \citep[e.g.,][]{agertz15,gatto17}. This regulation of star formation influences the number of galaxies of a given luminosity \citep[e.g.,][]{liu16,vogelsberger20,hutter21}. In addition to the hierarchical evolution of the dark matter halo mass function, baryonic processes—such as star formation efficiency and feedback mechanisms—can also evolve with redshift. In particular, stellar feedback is expected to have the greatest impact on low-mass galaxies, as it is more effective at expelling gas from their shallow gravitational potentials.

The number of galaxies per volume at a given ultraviolet (UV) luminosity, referred to as the UV luminosity function (LF), provides essential insights into the baryonic and non-baryonic processes that govern the growth of early galaxies. In particular, the shape and redshift evolution of the UVLF serve as powerful probes of the efficiency of star formation, the halo mass function, and feedback mechanisms. These processes collectively shape the integrated stellar mass assembly of the Universe. A direct comparison between the redshift evolution of the star formation rate density (SFRD) and the growth of dark matter halos offers critical information on the efficiency of conversion of baryons, namely how effectively gas is converted into stars within halos of varying masses and at different cosmic epochs \citep{dayal14,mashian16}.

For over two decades, the {\em Hubble Space Telescope} (\hst) has built a remarkable legacy in exploring the high-redshift Universe, discovering thousands of galaxies at redshifts greater than 6 and providing robust constraints on the overall shape of the UVLF during these epochs \citep{finkelstein15,bouwens15,atek18}. These results revealed a rapid steepening of the faint-end slope $\alpha$ of the UVLF and a systematic decrease in the number density of bright galaxies from $z=0$ to $z=9$, which marks the gradual build-up of the galaxy populations over time. These results were also consistent with the hierarchical growth of dark matter halos and indicated that the star formation efficiency remained relatively constant across cosmic time. However, reaching beyond a redshift of $z \approx 9$, and therefore stepping into the epoch when the first stars formed, had remained challenging due to \hst's limited infrared coverage. After two years of operations, the {\em James Webb Space Telescope} (\jwst) has identified hundreds of galaxy candidates at redshifts greater than 9, many of which are spectroscopically confirmed and located at record-breaking distances \citep{finkelstein22,fujimoto23,curtis-lake23,hainline23,arrabal23,roberts-borsani23,bogdan23,carniani24}. 

One of the main challenges arising from the first two years of JWST observations is an overabundance of UV-bright galaxies at redshifts beyond $z=9$ \citep[e.g.,][]{naidu22,castellano23,finkelstein23,austin23,adams24,chemerynska24} which contrasts with theoretical predictions \citep[e.g.,][]{mason15,crain15,tacchella18,dave19,yung19,dayal19}. The evolution of the UV luminosity density, \rhouv, and the corresponding star formation rate density, \rhosfr, is meanwhile significantly slower than anticipated by pre-\jwst\ models. This suggests a higher-than-expected production rate of UV photons at redshifts beyond $z=8$ \citep{naidu22,castellano23,chemerynska24}. This has now been observed in an increasingly large number of studies and independent fields, reducing potential field-to-field or environment-dependent variations \citep{donnan24,bouwens23,mcleod24,austin23,adams24}. Several physical scenarios are being explored to explain this slower than expected redshift evolution of the bright-end of the UVLF. These include (i) strong radiation-driven outflows that clear the dust content from star-forming regions in early galaxies \citep{Ferrara24}; effectively reducing the attenuation and increasing the UV luminosity of galaxies at all masses; (ii) the occurence of reduced or feedback-free starbursts (FFBs) \citep{dekel23}; producing an increase in star formation efficiency above a mass threshold that declines with redshift, leading to an increase in the luminosity of massive galaxies; (iii) a stochastic star formation history \citep[e.g.,][]{ciesla23,munoz23,pallottini23,sun2023,shen23,endsley24} which results in some low-mass galaxies contributing to the bright-end of the UVLF.

However, it remains unclear whether this slower evolution also extends to the faint-end of the UV luminosity function, where star formation efficiencies are expected to differ, or where stellar feedback might efficiently remove gas from the shallow gravitational potentials in these low-mass galaxies. Moreover, the physical motivations for ascribing the various scenarios have differing and testable, implications for the faint-end part of the LF. Surprisingly, this population of faint galaxies has remained largely unexplored in \jwst\ programs to date. This is primarily due to the programs' limited depth or their focus on targeting blank fields. Importantly, in standard $\Lambda$CDM, lower-mass halos are substantially more abundant than higher-mass halos, potentially implying that the bulk of early star formation might reside in this hitherto unexplored faint-end range of the UVLF. 

In this work, we report on the results of our analysis of the lensing-augmented observations of the \jwst\ GLIMPSE program (PIs: Atek \& Chisholm,~JWST-GO-3293). GLIMPSE obtained ultra-deep NIRCam observations of the $z = 0.35$ lensing cluster Abell S1063. Here, we present a comprehensive study of the faint-end of the galaxy UVLF and its evolution from $z\sim9$ to $z\sim15$. This work builds on a comprehensive characterization of the lensing effects, which permits us to probe the ultra-faint population of galaxies at these early epochs. We also derive the SFRD evolution across this redshift range integrating it for the first time down to the faintest magnitude reached thus far and explore its implications on galaxy formation models. This evolution is crucial for understanding the assembly of mass onto the very first DM halos, which seed large-scale structure formation in the Universe.

The paper is organized as follows: In Section \ref{sec:Observations}, we introduce the GLIMPSE survey and the associated \jwst\ observations, providing a detailed description of the data reduction process and photometric measurements. Section \ref{sec:Lensing} focuses on the mass model and the strong lensing constraints of Abell S1063. In Section \ref{sec:High-z}, we outline the methodology used to identify high-redshift galaxy candidates. Our approach to characterizing the effective survey parameters is described in Section \ref{sec:effective}, which includes completeness estimates (Section \ref{sec:completeness}) and the survey volume estimation (Section \ref{sec:volume}). The UV luminosity function results, along with comparisons to previous studies as well as theoretical predictions, are presented in Section \ref{sec:UV_LF}. The cosmic SFR density and its redshift evolution are discussed in detail in Section \ref{sec:SFRD}. Finally, Section \ref{sec:summary} provides a summary of our findings and conclusions of our work.

Throughout this work, we assume a flat $\Lambda$CDM cosmology with $H_0$ = 70 km s$^{-1}$ Mpc$^{-1}$, $\Omega_{M}$ = 0.3 and $\Omega_{\Lambda}$ = 0.7.

\section{The \jwst\ GLIMPSE program}
\label{sec:Observations}

This study utilizes ultra-deep NIRCam imaging of the lensing cluster Abell S1063 ($\alpha=$\ra{22}{48}{44.13},~ $\delta=$\decl{-44}{31}{57.50}) at a redshift of $z = 0.348$, obtained through the \jwst\ large program GLIMPSE (PIs Atek \& Chisholm). The observations include imaging data in seven broadband filters (F090W, F115W, F150W, F200W, F277W, F356W, and F444W) and two medium-band filters (F410M and F480M), achieving an unprecedented depth in the broadbands of approximately 30.8 mag at $5\sigma$ over the $0.8 - 5\mu$m wavelength range. Data reduction followed the procedure described in \citet{endsley24}, using the {\tt jwst\_1293.pmap} reference file, with further details provided in the GLIMPSE overview paper (Atek in prep.).

The reduction process included various enhancements beyond the standard Space Telescope Science Institute (STScI) pipeline, such as background subtraction, artifact correction, and cosmic ray rejection. To further improve the depth and access the immediate surroundings of the cluster field, we subtracted the bright cluster galaxies (bCGs) and intra-cluster light (ICL) from the images following methods outlined in \cite{shipley18, weaver23} and implemented in \cite{suess24}. Additionally, we incorporated ancillary \hst\ observations from the Hubble Frontier Fields \citep[HFF;][]{lotz17} and BUFFALO programs \citep{steinhardt20}. Prior to photometric measurements, we constructed empirical point spread function (PSF) models from stars in the field, convolving all \jwst\ and \hst\ images to the broadest PSF model, F480M in this case.

We performed source extraction on an inverse-variance weighted stack of the short-wavelength (SW) filters F090W+F115W+F150W+F200W and the long-wavelength (LW) filters F277W+F356W+F444W using {\sc SExtractor} \citep[][]{bertin96}. This process enables the SW bands to provide a factor of two improvement in spatial resolution compared to the LW bands. However, it may miss dusty or high-redshift galaxies that are detectable in the LW bands. To address this, we merged the SW and LW detection catalogs, retaining all sources from the SW catalog and including only those LW-detected sources whose central pixel falls within a region of the SW segmentation map with a value of zero, indicating no previously associated source. The photometry was computed within $D=0.2\arcsec$ apertures with {\sc Photutils}. Uncertainties on the photometry were calculated using the standard deviation within 2000 random apertures placed in adjacent source-free regions of the images.  To minimise the number of spurious objects, we excluded all sources located along diffraction spikes of bright stars, within $0.6\arcsec$ of the edge of the image, and within a 2\arcsec\ circle around a bCG. To perform the aperture corrections, we used the empirical PSF curves of growth to account for the fraction of the flux that falls outside our aperture size.

\section{Gravitational lensing}
\label{sec:Lensing}
To accurately constrain the high-redshift UVLF, it is crucial to have a robust gravitational lensing model of the foreground galaxy cluster lens in order to calculate the magnifications of the distant background sources and the survey volume. For GLIMPSE, we utilize a new parametric strong lensing (SL) model of AS1063, constructed with an updated version of the \citet{zitrin15a} analytic method \citep[see][]{furtak23a}. The model is fully analytic, in the sense that it is not limited to a prefixed grid resolution, which enables faster computation and higher resolution results. We model the cluster with two large-scale smooth dark matter (DM) halos, both parametrized as pseudo-isothermal elliptical mass distributions \citep[PIEMDs;][]{Kassiola93}. One halo is centered on the brightest cluster galaxy, while the other is positioned near a group of galaxies in the north-eastern region of the cluster \citep[e.g.,][]{Bergamini19,Beauchesne24}, slightly outside the field-of-view of GLIMPSE. Additionally, we model the 303 cluster member galaxies as dual pseudo-isothermal ellipsoids \citep[dPIEs;][]{eliasdottir07,Natarajan+97}. This relative simplicity of the mass distribution of AS1063 is the reason it was chosen by the GLIMPSE program, as this minimizes the lensing induced systematics typical for more complex SL clusters \citep[e.g.,][]{acebron2017,atek18,limousin2024}. The mass model is constrained with 75 multiple images, detected with \hst\ and MUSE, of 28 distinct sources, of which 24 have spectroscopic redshifts \citep[][]{balestra13,richard21,Beauchesne24,topping24}. We also use the parity information of a few arcs to further constrain the model. The model optimized a Monte-Carlo Markov Chain (MCMC) analysis and achieves a final lens-plane image reproduction RMS of $\Delta_{\mathrm{RMS}}=0.54\arcsec$. We refer the reader to Furtak et al. (in prep.) for more details on the mass modeling. A preliminary version of this model was previously also deployed in the analyses presented in \citet{topping24}, \citet{kokorev25}, and \citet{fujimoto25}.

We obtain magnification factors and their uncertainties for our sample by calculating the deflection at the position and photometric redshift of each object. The magnification uncertainties are drawn from the best-fit model's posterior distribution. In order to further account for typical SL systematics, conservatively, we add a systematic component of 15\,\% to the magnification uncertainties \citep[e.g.,][]{zitrin15a}. We use the model to verify if sources in our sample are multiply-imaged in section~\ref{sec:High-z}. We also construct the high-redshift source planes used for the volume computations in section~\ref{sec:completeness} and calculate the cumulative source plane areas as a function of magnification.

\section{High-redshift galaxy sample}
\label{sec:High-z}

\begin{figure*}
    \centering
        \includegraphics[width=0.45\linewidth]{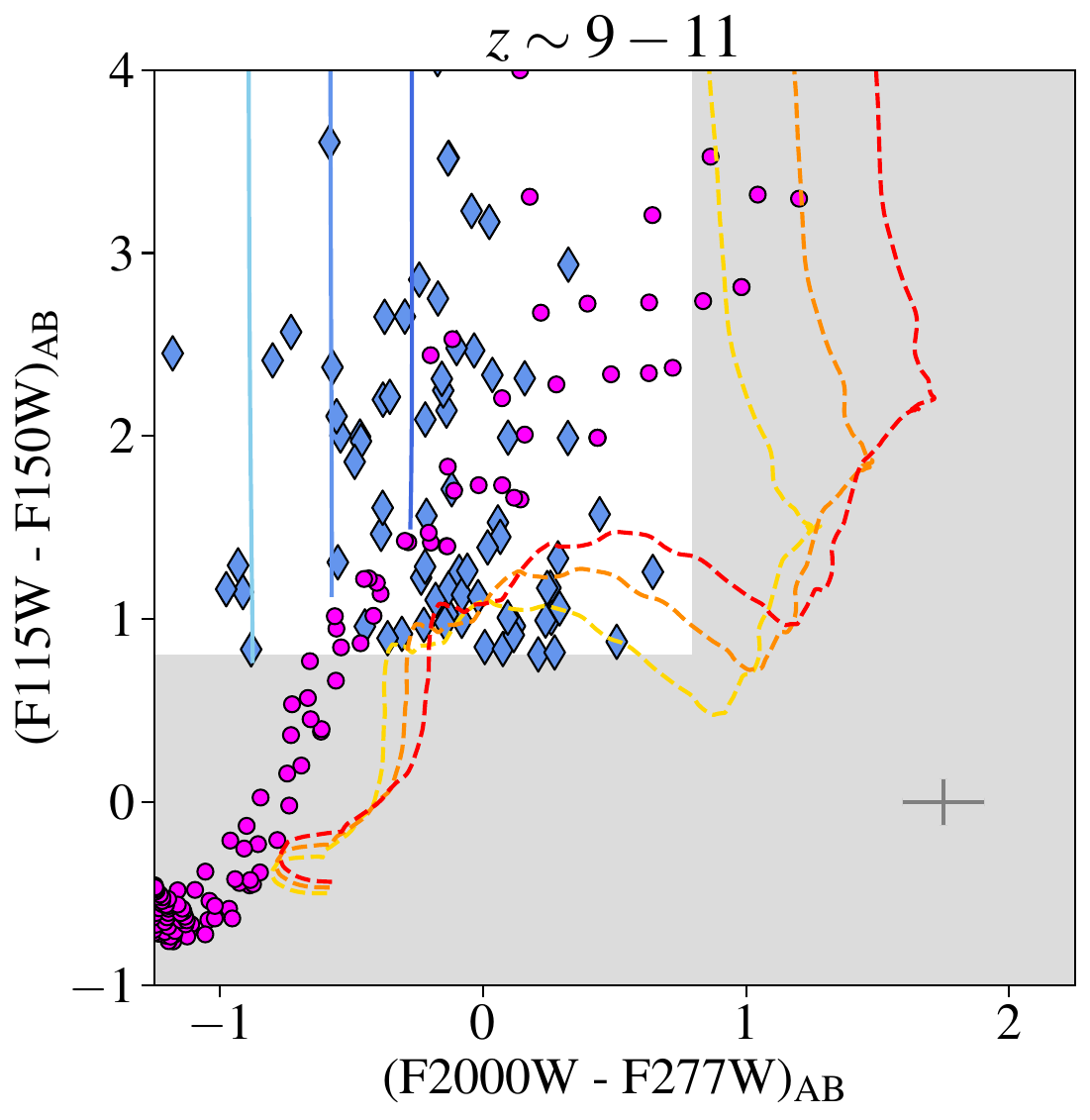}\includegraphics[width=0.45\linewidth]{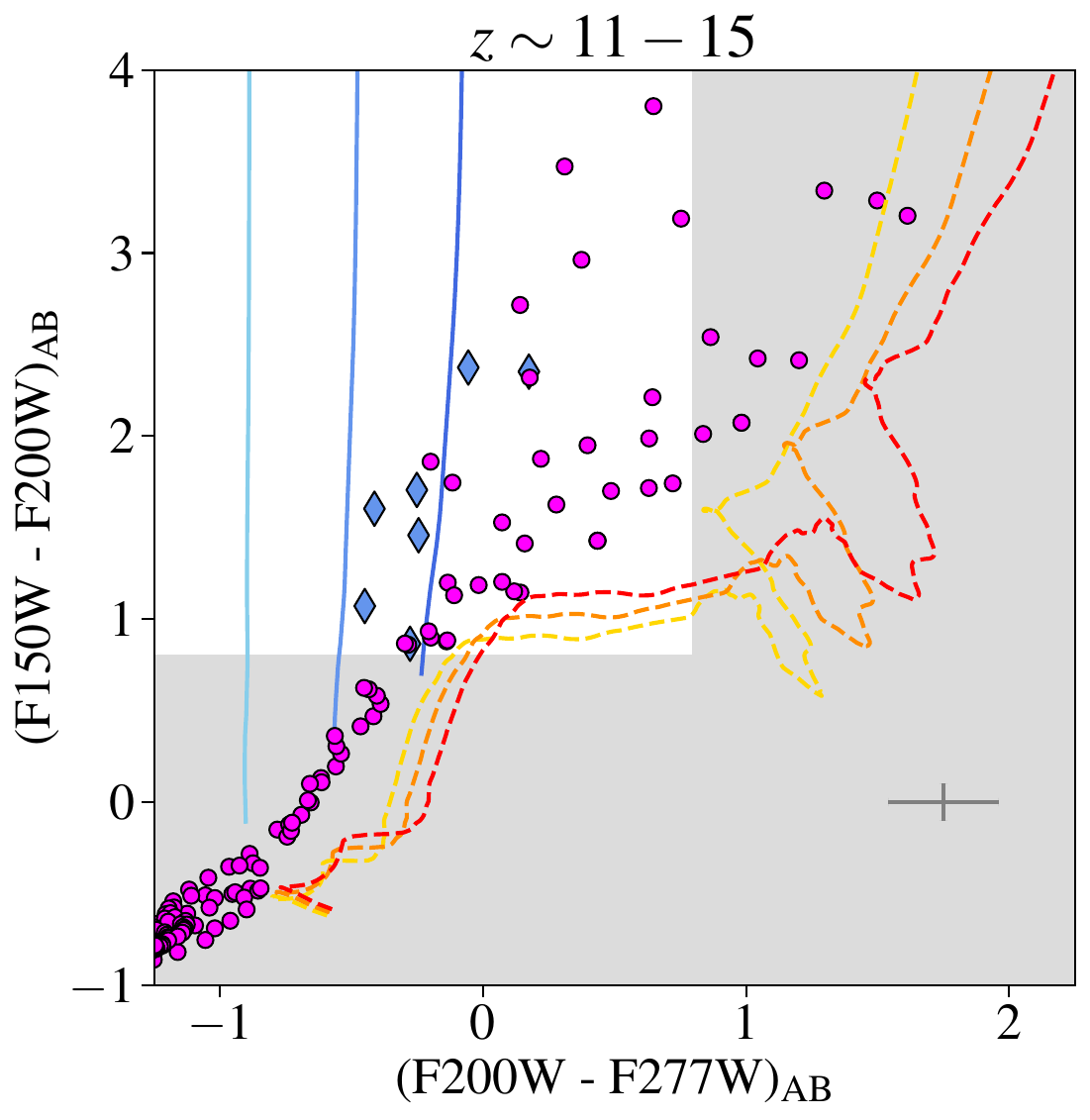}
        \caption{Color-color diagram for the high-redshift galaxy selection at $z \sim 9-11$ and $11-15$ (from left to right). The selected candidates are shown in blue diamonds with the selection window (white area) defined by the equations (\ref{eq:z9}) and (\ref{eq:z11}), respectively. The grey cross on the lower-right of each panel represents the average $1-\sigma$ uncertainty on the measured colors. The blue-solid lines are the expected color-color tracks of typical starburst galaxies at $z>9$ based on galaxy templates generated by \texttt{BEAGLE} (see Sect. \ref{sec:High-z} for details). The colored-dashed lines are color tracks of quiescent galaxies (potential low-redshift contaminants). Different values of attenuation were applied for high-z starburst and low-z quiescent templates, $A_v$ = [0, 0.25, 0.5] (yellow to red; blue to dark blue), assuming the SMC dust law. The magenta points show the color evolution of cold stars: brown dwarfs and M-class \citep[][]{Chabrier00,allard01}.}
    \label{fig:color-color_selection}
\end{figure*}

We begin by identifying high-redshift galaxy candidates using a color-color dropout selection, which relies on detecting the Lyman break caused by intergalactic medium (IGM) absorption \citep[e.g.,][]{steidel96,giavalisco02}. Galaxies in the redshift range $9<z<11$ dropout in the F115W filter, and galaxies between $z=11-15$ drop out in the F150W filter. 
To establish optimal high-redshift selection criteria, we computed color-color tracks for a range of galaxy templates and potential contaminating sources \citep[][]{atek23}. Using {\sc Beagle} \citep[][]{chevallard16}, we generated star-forming galaxy templates at redshifts 8 to 20
with dust attenuation values (A${\rm v}$) ranging between 0 and 0.5 mag, adopting the SMC extinction curve \citep[][]{pei92,reddy15}. To define a color space that minimizes contamination from low-$z$ interlopers, we also included quiescent galaxies simulated with {\sc GRASIL} \citep[][]{silva98} from the SWIRE library \citep[][]{polletta07}, with attenuation ranging from $A{\rm v}=1$ to 3 mags. Additionally, we incorporated brown dwarf interloper templates from \citet[][]{Chabrier00} and \citet{allard01}. Finally, we computed synthetic photometry across all \jwst\ bands and established Lyman break criteria for each dropout (redshift) sample.

Galaxies in the redshift range $9<z<11$ must satisfy the following color-color criteria:
\begin{equation}
	\begin{array}{l}
		m_{115}-m_{150}>0.8\\
	\land ~ m_{150}-m_{200}<0.8 \lor m_{200}-m_{277}<0.8
            
	\end{array}
  \label{eq:z9}
\end{equation}
and for $11<z<15$ galaxies, we adopt the following criteria:
\begin{equation}
	\begin{array}{l}
		m_{150}-m_{200}>0.8\\
		\land ~ m_{200}-m_{277}<0.8 \lor m_{277}-m_{356}<0.8
	\end{array}
 \label{eq:z11}
\end{equation}
where the first criterion picks the dropouts, while the other two criteria are used to exclude red contaminants. For each redshift range, we apply the following additional criteria to ensure the robustness of our sample. First, sources must be detected at the level higher than $3\sigma$ in bands red-ward of the break and at the 5$\sigma$ level in at least one band. We also exclude any source detected above the $2\sigma$ level in any band blueward of the break. When calculating the Lyman break, if a source's detection level falls below the $1\sigma$ magnitude limit in the dropout band, we substitute its magnitude with the $1\sigma$ uncertainty from the aperture measurement. Additionally, all sources undergo visual inspection to remove any remaining artifacts or spurious detections, such as residual diffraction spikes that were not flagged or contamination by a bright neighbouring object. To minimize contamination by dwarf stars, we also verify that our candidates are not point-source like and are not too compact. Figure \ref{fig:color-color_selection} illustrates the process of dropout selection and the color-color window adopted for each redshift range. The galaxy candidates are shown in the color-color diagram alongside the main sources of contamination. 

\begin{figure*}
    \centering
    \includegraphics[width=\linewidth]{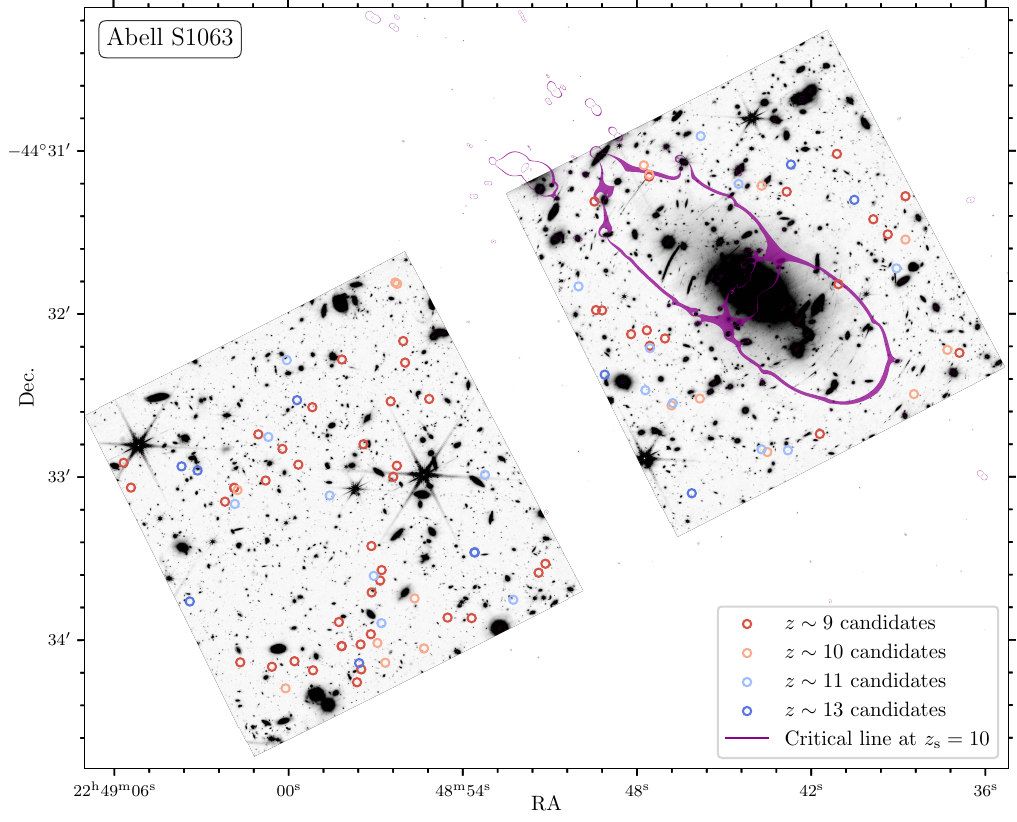}
    \caption{GLIMPSE NIRCam image of AS1063 in the F444W-band showing the positions of our high-redshift candidates in each redshift bin. The critical line for a source at $z_{\mathrm{s}}=10$ from our lens model (see Section~\ref{sec:Lensing}) is overlaid in purple.}
    \label{fig:image}
\end{figure*}

\begin{figure}
    \centering
    \includegraphics[width=\linewidth]{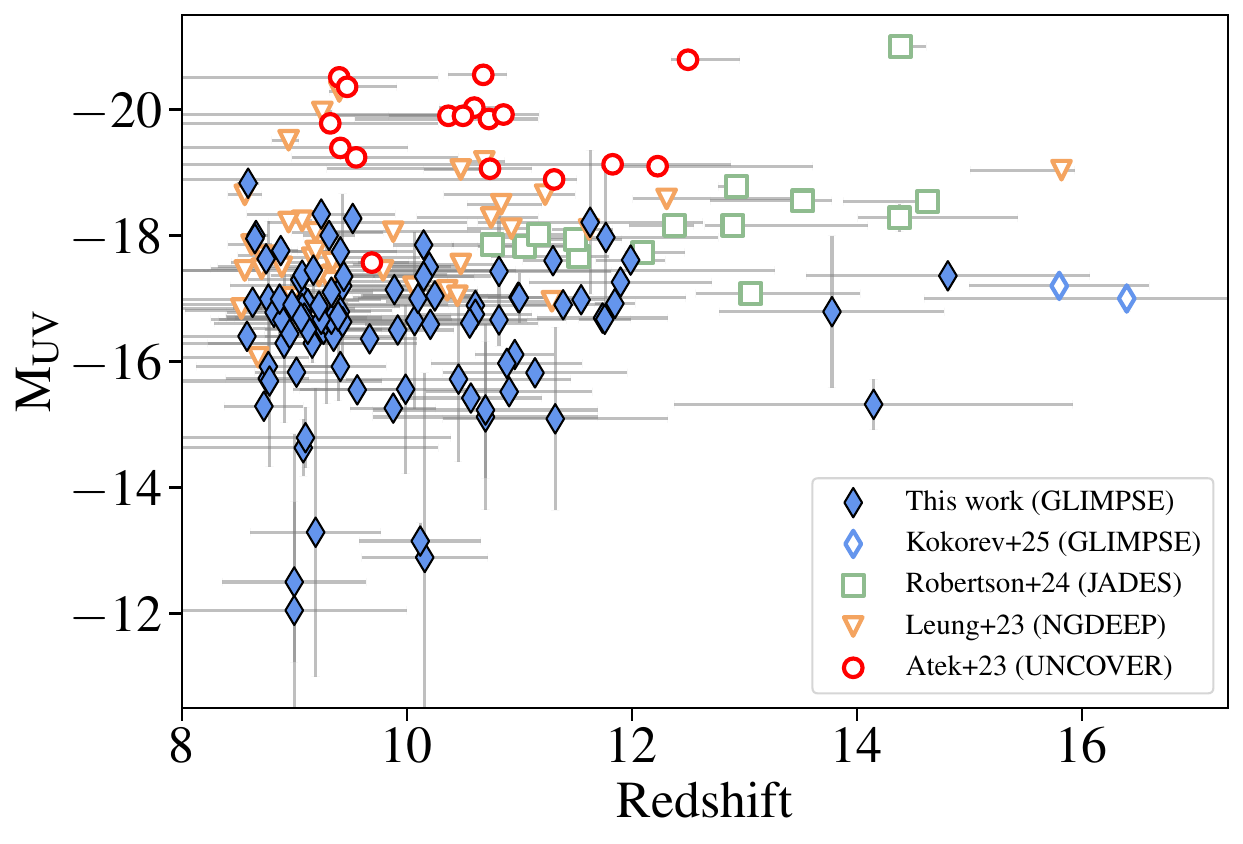}
    \caption{Intrinsic absolute magnitude of the galaxy sample as a function of redshift. The sample from this work is presented in the blue diamonds, together with 2 galaxy candidates from \citet{kokorev25} (empty blue diamonds). The uncertainties represent photometric errors along with the propagated magnification uncertainties. For comparison, we also show results from the deepest \jwst\ surveys that probe similar redshift range: JADES \citep[][empty green square]{Robertson24}, NGDEEP \citep[][empty orange triangle]{leung23} and the lensed survey UNCOVER \citep[][empty red circle]{Atek_2023b}.}
    \label{fig:muv_z}
\end{figure}

In parallel, we computed photometric redshifts for the entire catalog of sources using the Python implementation of {\sc Eazy} \citep{Brammer2008}. The analysis incorporated the \jwst\ GLIMPSE photometry, utilizing aperture-corrected $D=0.2''$ \textit{JWST} flux densities. We adopted the {\tt blue\_sfhz\_13} template library, which includes redshift-dependent star formation histories based on high-$z$ results from recent \jwst\ observations \citep[][]{carnall23}, as well as dust attenuation \citep{Calzetti00} and IGM attenuation \citep{asada24}. The redshift grid spanned the full range, from $z=0.01$ to $z = 30$. In the end, all the dropout-selected sources have best-fit photometric redshifts within the selection window. Hereafter, we adopt the {\sc Eazy} best-fit solution as the photometric redshift, along with the corresponding uncertainty derived from the redshift probability distribution. The final galaxy sample contains 98 galaxies at $z\sim9-11$ and 7 galaxies at $z\sim11-15$. 
The positions of the galaxy candidates in the GLIMPSE field are illustrated in Figure \ref{fig:image} and the full list of the sample is provided in Table \ref{tab:z_9_11}. Figure \ref{fig:muv_z} presents the intrinsic (de-lensed) UV magnitude of the candidates as a function of redshift. This figure demonstrates how GLIMPSE significantly reveals fainter galaxies across all redshifts compared to previous \jwst\ surveys. With deep observations and gravitational lensing, we pushed the observational limit down by $\sim3$ magnitudes in \muv, allowing us to access an important and hitherto unexplored regime of faint galaxies.

\section{Effective Survey Volume Estimates}
\label{sec:effective}

\subsection{Completeness Estimates}
\label{sec:completeness}

Before computing the UV luminosity function, we first need to accurately estimate the completeness function of our survey and its associated uncertainties. This process follows the source-plane method described in \citet[][]{atek18} and \citet[][]{chemerynska24}. The core principle involves implementing mock galaxies, spanning a wide range of properties and redshifts, directly in the source plane reconstructed from the survey image using our cluster mass model. The primary advantage of this approach is the ability to account for all lensing effects simultaneously, including flux magnification, distortion, geometric deflection, multiple imaging, and the resulting variations in incompleteness due to lensing.

For this procedure, we generated 150,000 mock galaxies at $z\sim9-12$ and 50,000 across the remaining redshift ranges. These mock galaxies were randomly distributed with redshifts within our redshift range and intrinsic absolute magnitudes ranging from $-20$ to $-10$ mag. To ensure realistic source sizes, we applied the size-luminosity relation derived by \cite{yang2022} for galaxies fainter than $-16.5$ mag and by \cite{shibuya15} for brighter sources. The sources were placed randomly in the source plane, with a higher density in regions of high magnification, to improve the statistical reliability of the completeness function. This adjustment compensates for the tendency of high-magnification regions to be undersampled.

\begin{figure}
    \includegraphics[width=0.9\linewidth]{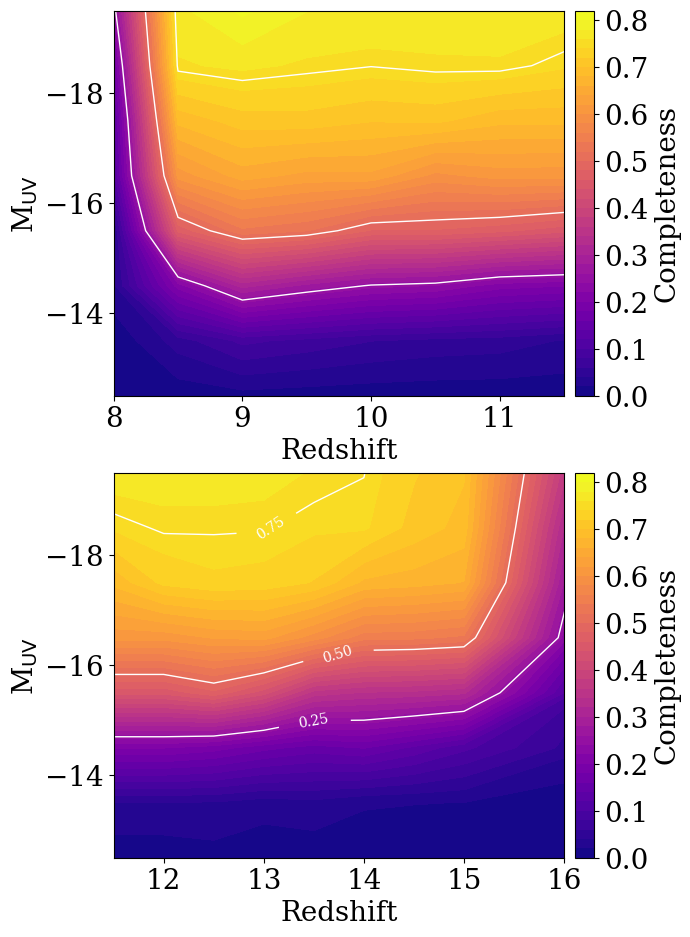}
    \caption{ The survey completeness of each redshift range based on the color-color selection (see Section \ref{sec:completeness}). Each 2D map shows the completeness fraction (color bar) as a function of both the intrinsic magnitude \muv\ and the redshift. The completeness reaches a maximum of around 80\% near magnitudes \muv$>-18$ mag and at the center of the redshift selection function. The lowest levels of completeness are for galaxies fainter than \muv $= -13$ mag, which have less than 10\% completeness.
    }
    \label{fig:completeness}
\end{figure}

We follow the procedure outlined in \cite{chemerynska24}, utilizing SED models generated with {\tt BEAGLE} \citep{chevallard16} to compute synthetic fluxes for the mock galaxies. The templates are based on stellar population models from \citet[][]{bc03} assuming constant star formation, an ionization parameter log($U$) in the range (-4, -1), an SMC extinction law \citep{pei92}, and a constant metallicity of $Z= 0.1 Z_{\odot}$, with an extinction in the range $A_{\rm V}=0-1.25$ applied. The templates were redshifted and normalized to match the observed mock magnitudes in their detection filters (F150W, F200W, F277W), corresponding to the rest-frame UV. Images of the simulated galaxies were then created using {\tt GalSim} \citep{rowe2015galsim}, based on their physical properties. These sources were injected into the actual \jwst\ bCG subtracted images of the galaxy cluster, placing 100 galaxies at a time to prevent source overlap. Finally, we perform the same photometric measurements and apply the same color-color selection to the full sample of mock galaxies. It is important to note that when multiple images of a mock galaxy are detected, we retain only the brightest candidate. The final completeness function is then computed by comparing the output catalog to the original input catalog, as a function of intrinsic magnitude and redshift. Figure \ref{fig:completeness} presents the results of our completeness estimates across different redshift ranges. It illustrates the evolution of completeness levels, reaching a maximum of approximately 80\%, as a function of intrinsic (unlensed) UV magnitudes and redshifts. For the faintest sources, which require high amplification factors, the completeness drops to just a few percent, primarily due to the limited surface area with high magnification.

\begin{figure}
    \centering
    \includegraphics[width=\linewidth]{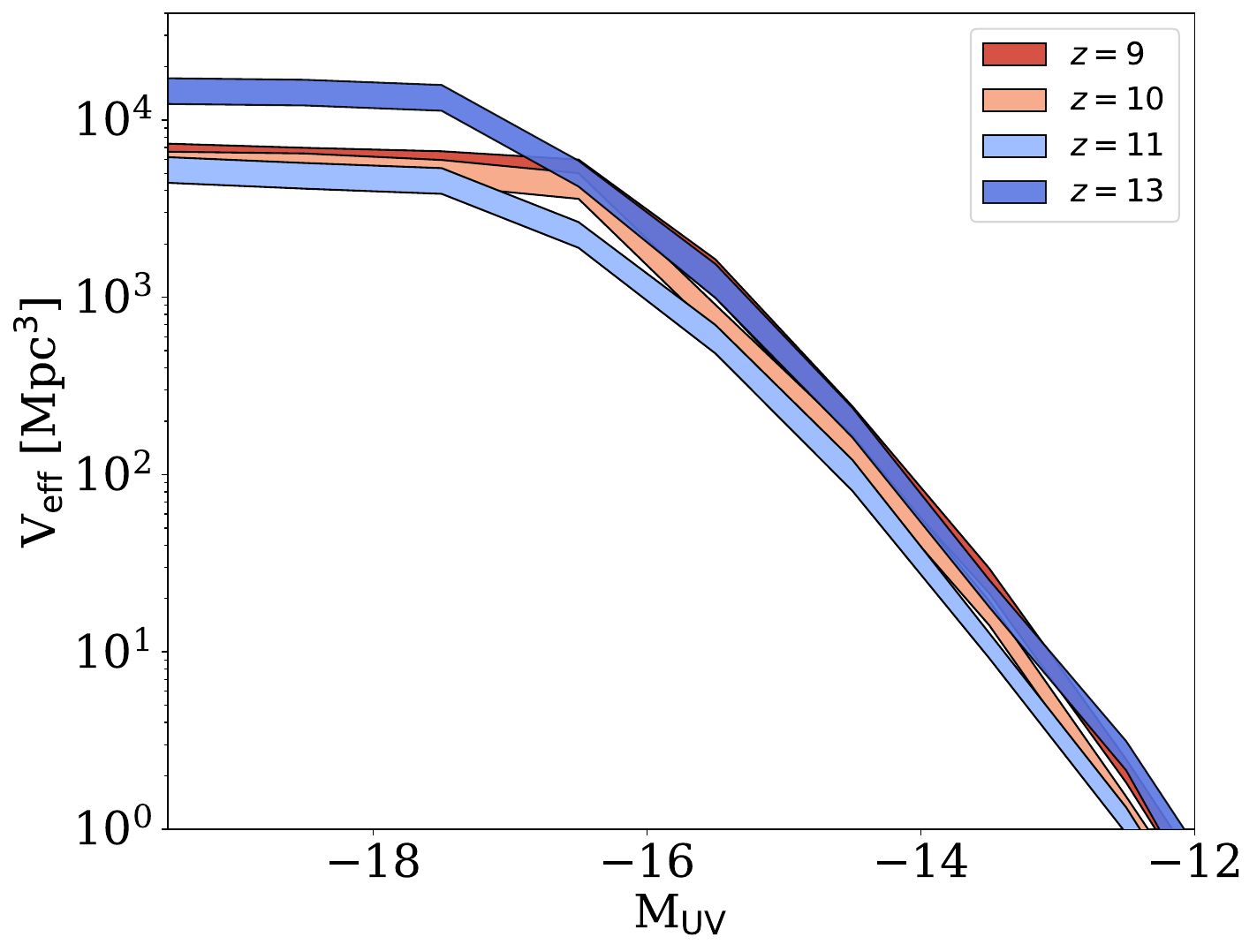}
    \caption{ The effective GLIMPSE survey volume through the AS1063 lens as a function of the intrinsic (de-lensed) absolute magnitude. Each curve represents the cumulative volume estimate at a given redshift and \muv, based on the lensing-corrected area, the redshift selection function, and the completeness function (cf. Figure \ref{fig:completeness}). The total survey volume ranges from $\approx 8\times10^{3}$ Mpc$^{3}$ at $z=9$ to $19\times10^{3}$ Mpc$^{3}$ at $z=13$}.
    
    \label{fig:volume}
\end{figure}

\subsection{Survey Volume Estimates}
\label{sec:volume}

The effective volume is determined by the comoving maximal volume, which is distorted by gravitational lensing, multiplied by the completeness function \citep{atek18}. The comoving volume depends on the area associated with a specific magnification at a given redshift. We calculate the surface area corresponding to the magnification at a given redshift, as well as the minimal magnification required for a galaxy of a certain magnitude to be detected. The surface area is determined by summing the cumulative area over regions where the magnification exceeds the minimal magnification. The total area of the cluster primarily establishes the maximum volume, as the surface area becomes significantly smaller at progressively higher magnifications. The resulting effective volume for each redshift bin over the intrinsic absolute magnitude is shown in Figure \ref{fig:volume}.

In addition to the uncertainties in the effective volume, we account for other sources of uncertainty when calculating the galaxy number density in each magnitude bin of the luminosity function. First, we estimate the Poisson confidence intervals following \cite{Gehrels86}, which accounts for small-number statistics. Second, we compute the fractional error due to cosmic variance for each magnitude bin and redshift, using the corresponding lensing-corrected volume. To estimate cosmic variance, we employ the calculator from \cite{Trapp20}, resulting in CV = 34 -- 50\%. Finally, we account for all uncertainties affecting the intrinsic magnitudes of galaxies, including those arising from magnification, photometric measurements, and photometric redshift estimates. To incorporate all these uncertainties into our binning scheme, we use Monte Carlo simulations that allow galaxies to shift between bins based on their uncertainty distributions. This approach has the advantage of robustly capturing variations due to the arbitrary choice of bin positions and sizes.

\section{The Ultraviolet Luminosity Function}
\label{sec:UV_LF}
After selecting the high-z candidates, we are now ready to compute the UV luminosity function. In order to optimize the number of galaxies in each bin at each redshift, we divide the sample into four redshift sub-samples: $z =$ 8.5 -- 9.5, 9.5 -- 10.5, 10.5 -- 11.5 and 11.5 -- 15 (centered on $z =$ 9, 10, 11, 13; 
see Tab. \ref{tab: UV_LF}). To compute the UVLF, we additionally adopt magnitude bins ranging from $-18.5$ mag to $-12.5$, with a bin width of $\Delta$ mag = 1. The UV luminosity function at each redshift bin is computed as follows:
\begin{equation}
    \centering
        \phi(M_i)dM_i=\frac{ N_{{\rm obj},i}  }{{V_{\rm eff}}(M_i)}
\end{equation}
where $N_{{\rm obj},i}$ represents the number of galaxies in each magnitude bin and $V_{\rm eff}(M_i)$ denotes the effective volume in the $i^{\rm th}$ bin of absolute UV magnitude $M_i$.

\begin{figure*}
    \centering
        \includegraphics[width=0.48\linewidth]{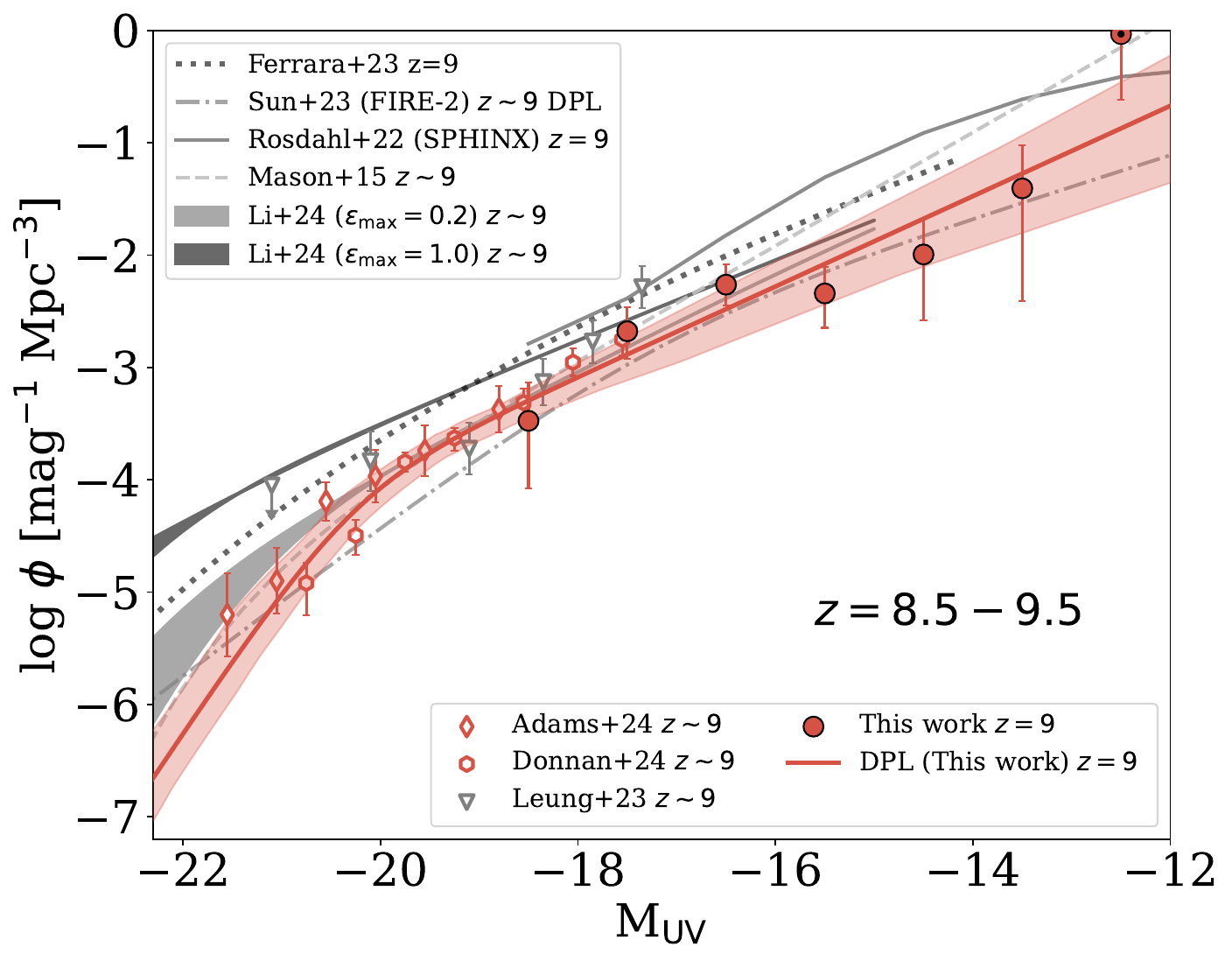}
        \includegraphics[width=0.48\linewidth]{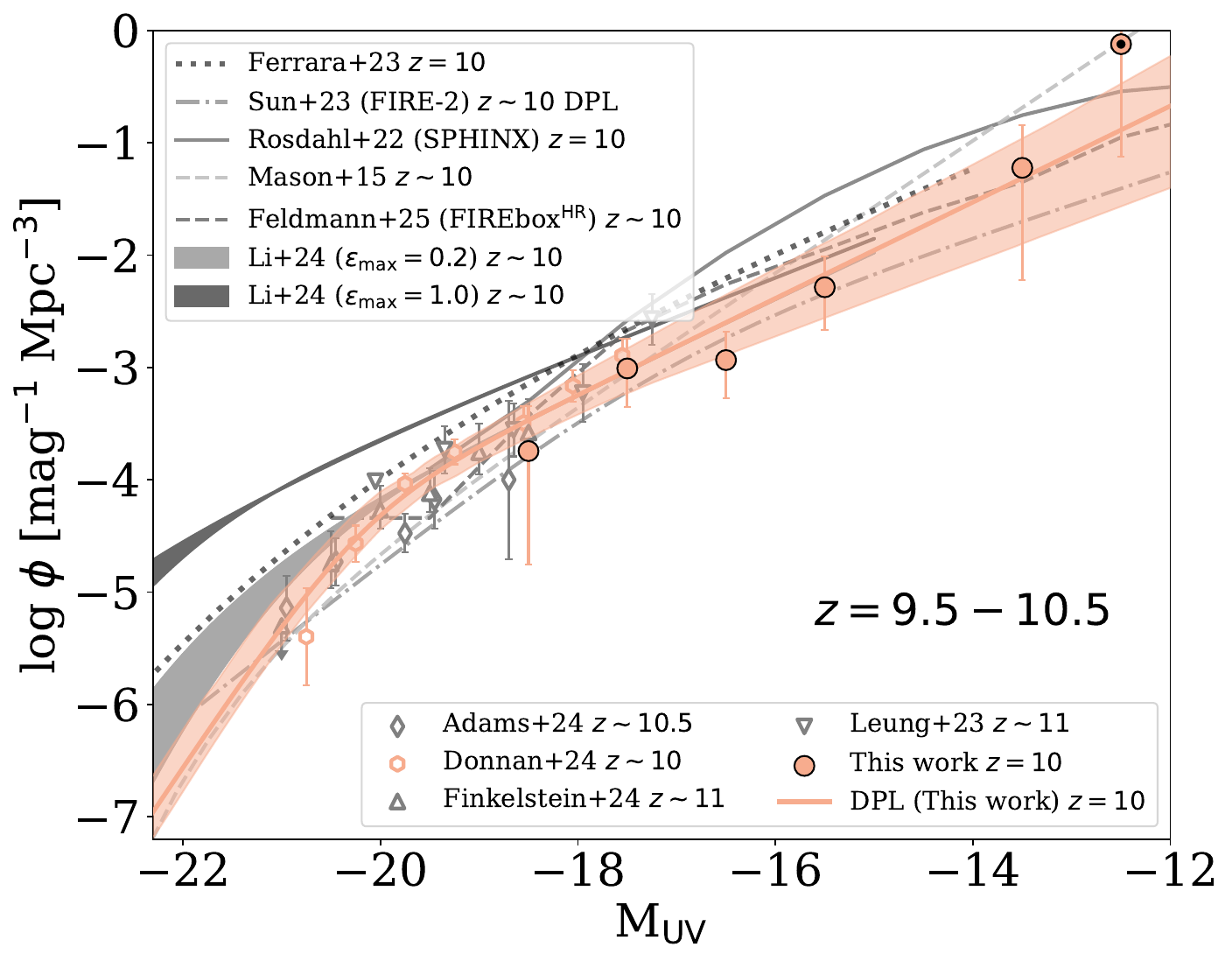}
        \includegraphics[width=0.48\linewidth]{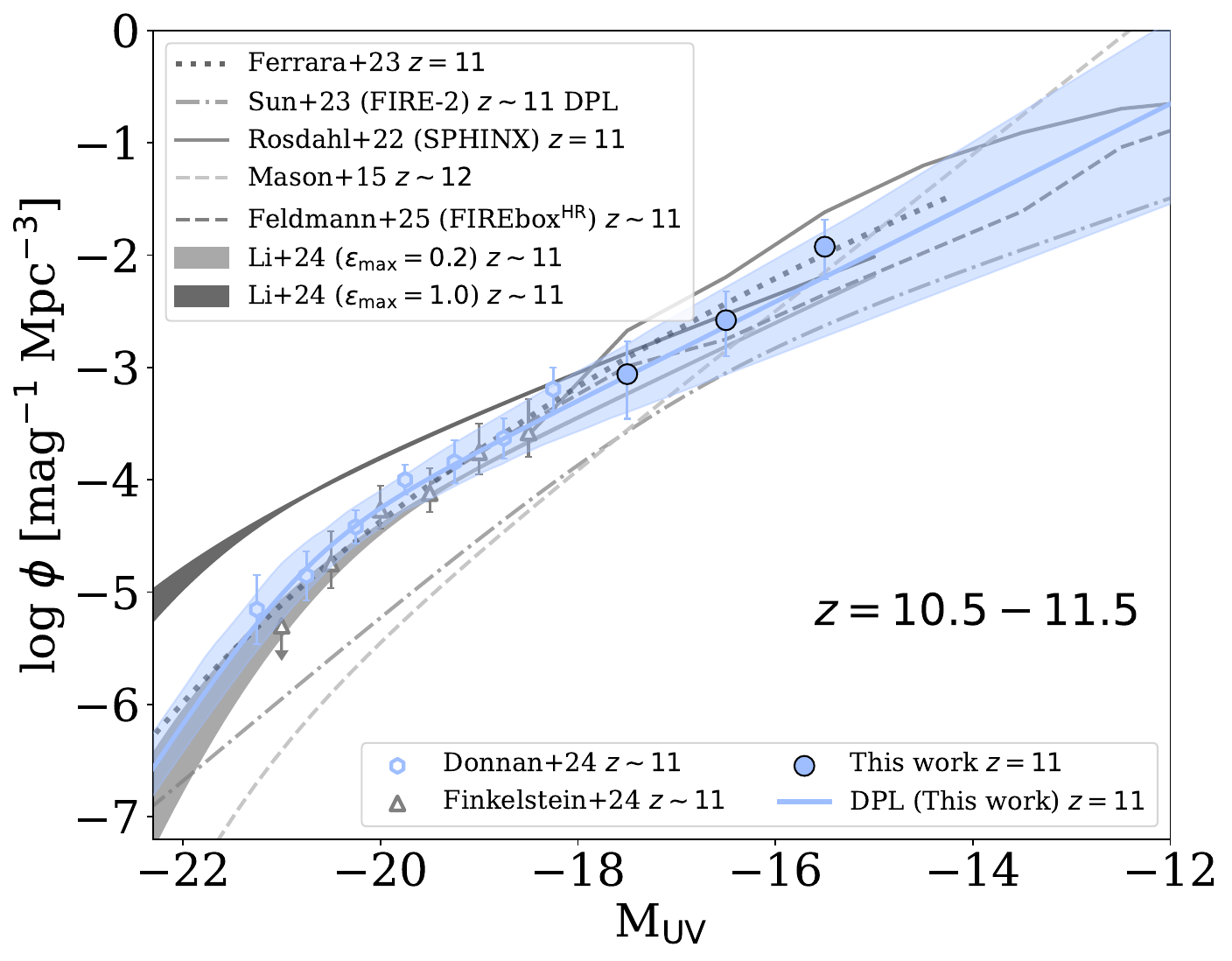}
        \includegraphics[width=0.48\linewidth]{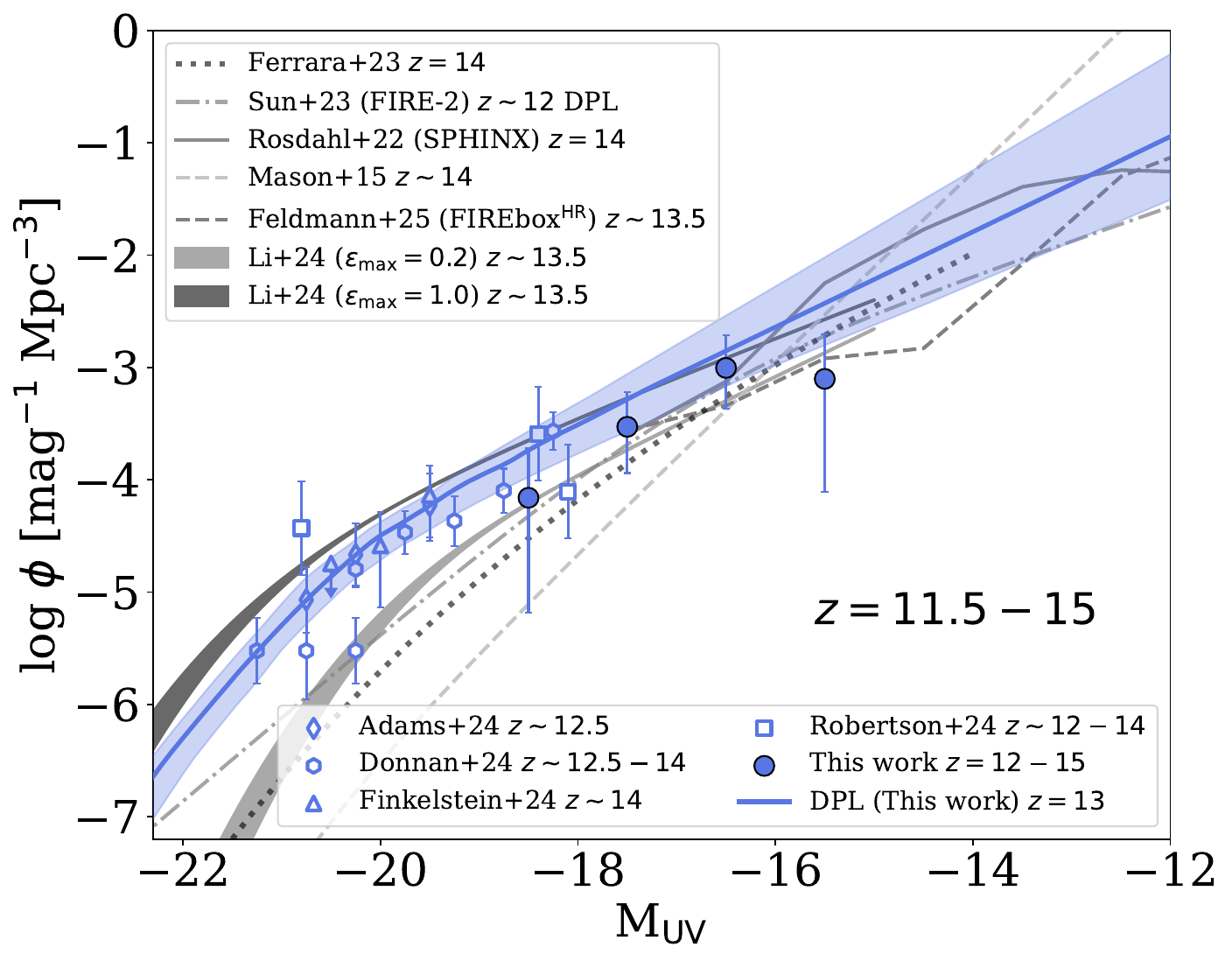}
        \caption{\jwst's GLIMPSE of the ultra faint-end of the UV Luminosity function at $z>9$. Each panel presents our UVLF determination at a given redshift, alongside literature results at the bright-end. {\bf Top-left:} The UVLF at $z=9$ shown as red circles with associated $1\sigma$ uncertainties (the dark circle denotes the faintest bin, where the completeness is less than $\sim 10\%$). Literature results are shown with various grey symbols (cf. legend) from \citet[][]{adams24,donnan24,leung23}. The red line represents the best-fit double power law relation, which includes our results at the faint-end and the data points of \citet[][]{donnan24} and \citet[][]{adams24} at the bright end. The red shaded region denotes the $1\sigma$ uncertainty of the best-fit. {\bf Top-right} same as the previous panel but for $z=10$ and GLIMPSE results are shown in orange, including data from \citet[][]{donnan24} at the bright-end. {\bf Bottom-left} same as previous but for $z=11$. Our results are shown in light blue. {\bf Bottom-right} same as previous panels, but for the redshift range $z=13$. Here, the fit is performed on data points from various literature results at the bright end \citep[][]{Robertson24,finkelstein23,adams24,donnan24} in addition to our GLIMPSE data points at the faint-end. Several theoretical determinations are represented in each of the panels and include the SPHINX simulations \citep[][]{rosdahl22}, the FIRE-2 simulations \citep[][]{sun2023}, the FIREbox$^{\mathrm{HR}}$ simulations \citep[][]{Feldmann25}, models from \citet[][]{ferrara23}, \citet[][]{mason15} and \citet[][]{li23}. 
       }
    \label{fig:UVLF}
\end{figure*}

\begin{figure*}
    \centering
    \includegraphics[width=0.8\linewidth]{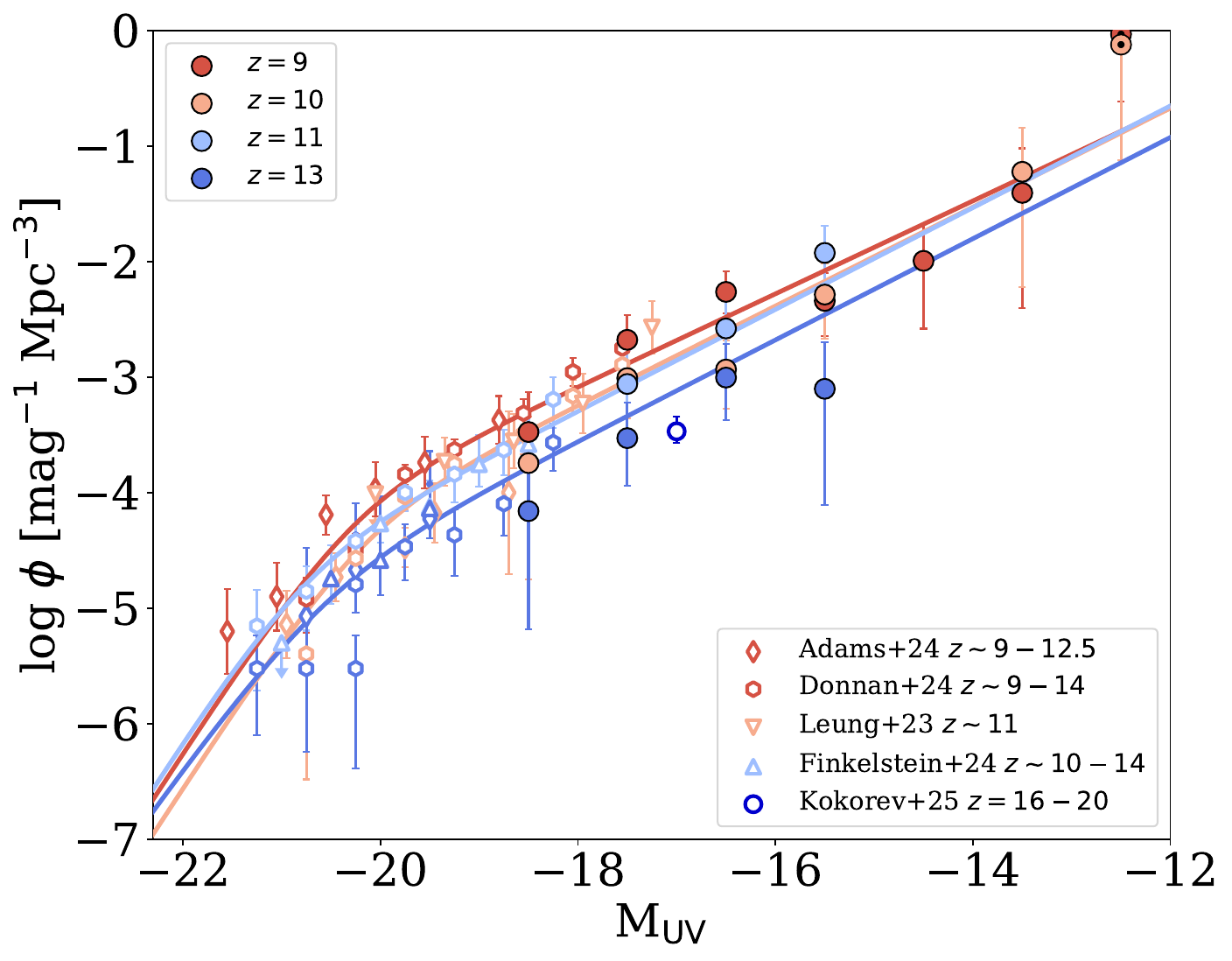}
    
    \caption{The evolution of the faint-end of the UV luminosity function through Cosmic Dawn. For each redshift, our UVLF points are shown with circles, while the best-fit DPL relation is represented with a solid line. For each redshift, literature results are shown with different symbols, but with the same color as our data points. 
    }
    \label{fig:UVLF_all}
\end{figure*}

\subsection{The ultra-faint-end of the UV luminosity function at $z>9$}

Our derived UVLF at redshifts above $z=9$ extends previous estimations into a previously unexplored regime, where we are also able to constrain the faint-end of the luminosity function for the very first time. These constraints are critical for understanding the abundance and properties of faint galaxies in place in this early cosmic epoch. Furthermore, they provide valuable insights into the processes governing galaxy formation and evolution in the first billion years of cosmic history.

In Figure \ref{fig:UVLF}, we plot the UVLF determinations at each redshift interval, and provide the best-fit relation at each redshift, including previously published results for the bright end. The derived UVLF data points are provided in Table \ref{tab: UV_LF}. 

Recently, numerous studies have investigated the UVLF at $z>8$ using \textit{JWST} observations \citep[e.g.,][]{adams24,bouwens23,casey23,castellano23,chemerynska24,donnan24,finkelstein23,Harikane2024,leung23,Perez-Gonzalez23,willott24}. Most of these studies have provided robust constraints on the bright end of the LF at high-redshift, revealing that it is better described by a double power law (DPL) rather than the classical Schechter function, since the latter underestimates the numbers of observed bright galaxies. This important result represents one of the most unexpected findings from the initial years of \textit{JWST} operations: that the high number density of observed UV-bright galaxies at  $z>9$ exceeds pre-\jwst\ theoretical predictions.
We fit our results for the UVLF with a double power law function:
\begin{equation}
    \centering
    \phi(M) = \dfrac{\phi^*}{10^{0.4(M-M^*)(\alpha+1)}+10^{0.4(M-M^*)(\beta+1)}}
\end{equation}
where $\phi^*$ is the number density normalization of the LF, $M^*$ is the characteristic magnitude, that separates the bright and faint-end distributions, $\alpha$ is the faint-end slope, and $\beta$ is the bright-end slope. Our sample primarily consists of faint objects (\muv$>-18.5$), enabling us to accurately constrain the faint-end of the luminosity function. 

Since our survey primarily probes the faint regime and a small volume, we rely on bright-end data points from the literature that analyzed wide-area survey data. In our fitting procedure, we incorporate results from the \cite{adams24} and \cite{donnan24} studies, which derive the UVLF over the same redshift range as our work using a compilation of public surveys. Given the significant dispersion in the bright-end UVLF measurements at $z =$ 11.5 -- 15 amongst different studies, we include multiple datasets in our fit for this redshift range from \citet{adams24,donnan24,finkelstein23,Robertson24}. Furthermore, to ensure the robustness of our faint-end fit, we exclude the lowest luminosity bin ($-12.5$ mag) due to its limited completeness ($\sim 10\%$).

To determine the best-fit parameters and their uncertainties, we perform Markov Chain Monte Carlo (MCMC) simulations. The characteristic magnitude ($M^{*}$), the faint-end slope $\alpha$, the bright-end slope $\beta$,  and the number density normalization $\phi^*$ are left as free parameters in all redshift bins. All the values of the best-fit parameters for the DPL function at each redshift are summarized in Table \ref{tab:param}. 

\begin{figure*}
    \centering
        \includegraphics[width=0.99\linewidth]{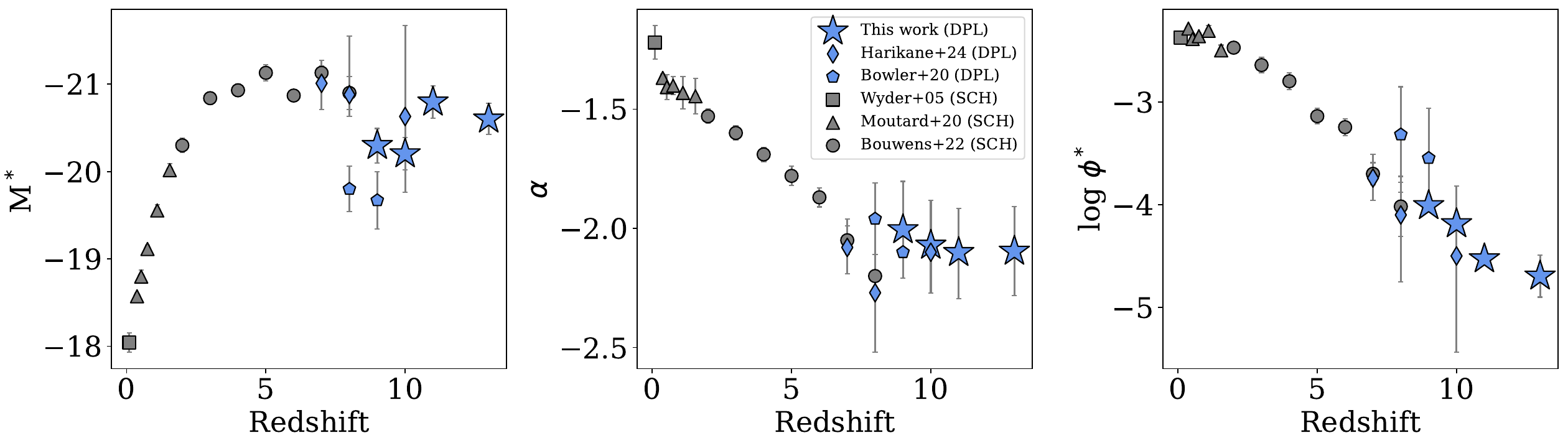}
        \caption{The evolution of the characteristic magnitude ($M^*$), the faint-end slope ($\alpha$) and the normalization ($\phi^*$) with redshift. Our results, which are based on a DPL parametrization of the LF, are presented as blue stars. In contrast, the grey squares, triangles and circles represent constraints from previous studies at $z = 0.1 -8 $: \citet{wyder05}, \citet[][]{Moutard20}, and \citet{bouwens22c}, respectively, which were all derived assuming a Schechter function (SCH). We also include results of DPL parametrization for the LF  from \citet[][blue pentagons]{bowler20} at $z=8$ and $z=9$, and \citet[][blue diamonds]{Harikane2024} at $z = 7-10$.}
    \label{fig:param}
\end{figure*}

\begin{figure}
    \centering
        \includegraphics[width=0.8\linewidth]{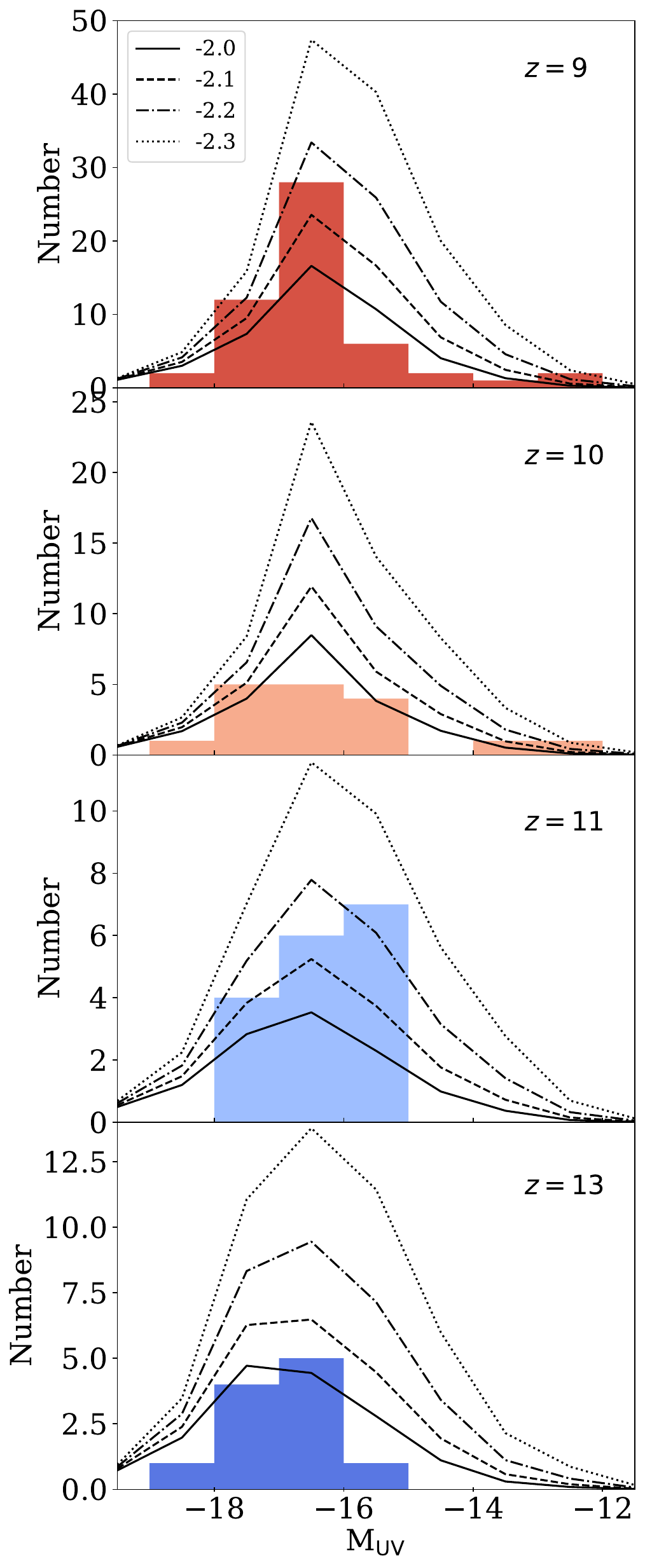}
        \caption{The number of sources as a function of magnitude at redshift from 9 to 15 
        (from top to bottom). The solid, dashed, dot-dashed, and dotted lines represent the predicted number of sources per bin for different values of the faint-end slope $\alpha$. Results show that very steep faint-end values tend to overestimate the number of sources, while $\alpha$ values in the range of $-2.0$ to $-2.3$ provide good constraints on the faint-end of the UVLF.}
    \label{fig:faint-end}
\end{figure}

The overall shape of the luminosity function, combining our results with previous bright-end determinations, is well characterized by a DPL function. Thanks to the combination of ultra-deep imaging and gravitational lensing, our study pushes the luminosity frontier to extremely faint magnitudes. This allows us to provide the first estimates on the very faint-end of the high-redshift UVLF, down to \muv $\approx -14$, on average. The best-fit UVLF parameters using GLIMPSE data at various redshifts are presented in Table \ref{tab:param}. We observe a slow decrease in $\phi^*$ as a function of redshift, from log($\phi^{*}$/Mpc$^{-3}$ mag$^{-1}$)$=-4.01\pm 0.04$ at $z=8.5-9.5$ to log($\phi^{*}$/Mpc$^{-3}$ mag$^{-1}$)$=-4.70\pm 0.20$ at $z=11.5-15$. This slow evolution of the GLIMPSE data directly mimics the recent \jwst\ findings \citep[][]{finkelstein23,adams24, willott24,donnan24}, which reported a more modest evolution of the bright end of the UVLF between $z=8-15$. Figure \ref{fig:UVLF_all} combines the UVLFs derived at different redshifts. While the LF normalization decreases slightly with increasing redshift, the faint-end slopes remain nearly constant from $\alpha=-2.01\pm 0.20$ at $z=8.5-9.5$ to $\alpha=-2.10\pm 0.19$ at $z=11.5-15$. This absence of strong evolution is also consistent with the recent findings of \citet[][]{Perez-Gonzalez23} wherein using deep NIRCam observations of the parallel field of the MIDIS survey \citep[][]{Ostlin24}, they investigate the UVLF across \muv$ =-20$ to $-17$ mag at $z=9-12$. The faint-end slope varies from $\alpha=-2.30^{+0.16}_{-0.24}$ at $z=9$ to $\alpha=-2.19^{+0.26}_{-0.39}$, consistent within 1$\sigma$ uncertainties with our derived values, and also indicating slow evolution. While surveys such as CEERS and PRIMER primarily focused on the bright end of the luminosity distribution at these redshifts, deeper surveys like the JADES Origins Field (JOF) and NGDEEP have used significantly longer integration times to probe the population of fainter galaxies. In particular, \citet[][]{leung23} present a DPL fit to the UVLF measured at $z=8.5-12$ in the NGDEEP field, finding a slow evolution in the LF from $z=9$ to $z=11$ and that the faint-end slope varies from $\alpha=-2.5\pm0.4$ to $\alpha=-2.2\pm0.2$, which is statistically consistent with no variation. Using observations from the deepest JADES field, \citet[][]{Robertson24} derived the UVLF at $z=12-14$, adopting a Schechter function form, and found that the LF normalization, $\phi^{*}$, decreased by a factor of $\approx 2.5$ moving from $z=12$ to $z=14$, which is more pronounced than the results of \citet{leung23}. Additionally, in the same field, \cite{Whitler25} explored the UVLF at $z>9$ for UV luminosities of $ -21<$ \muv $<-17$ mag. They utilize both the Schechter function and DPL fitting function forms, determining a steep faint-end slope of $-2.5 < \alpha < -2.3$ at redshifts $z = 9-16$.  

In Figure \ref{fig:param} we summarize the evolution of $\alpha, M_{\star}$ and $\phi^*$ over a wide redshift range by comparing our results with a compilation of results from the literature at lower redshifts \citep[][]{wyder05,Moutard20,bouwens22c}. Consistent with previous studies, the characteristic magnitude, $M^*$, plateaus at approximately $-20.7$ mag. This value is somewhat lower than earlier estimates for $z = 5-9$, primarily due to differences in parametrization between the Schechter function and the DPL function. Notably, \citet[][]{bowler20}, assume a DPL function to model the luminosity function, also report lower values for $M^*$. Similar trends are observed for the faint-end slope, which evolves rapidly up to $z\sim 8$, and stabilizes around $\alpha \sim -2.1$ at $z > 9$. Lastly, while there is a pronounced evolution in $\phi^*$, we do not detect an ``acceleration'' for $z > 10$. Such acceleration has been suggested to align with the evolution of the underlying dark matter halo function, which would indicate a constant star formation efficiency in these early galaxies \citep[][]{mason15,oesch18,tacchella18,Sabti:2021xvh}. 

We also examined how the predicted number of galaxies changes with varying faint-end slopes \cite{bouwens22c}. In Figure \ref{fig:faint-end}, we present the magnitude distribution of our sample at each redshift, together with the predicted values of the faint-end of UVLF. This visually shows what faint-end slope reproduces best the observed distribution. The faint-end slope appears to be between $-2.0$ and $-2.2$, which is consistent with our findings from the UVLF. Given that our fit is based on our data as well as other studies, it is clear that slight variations between observed numbers and those predicted by the faint-end slope can occur.
This figure indicates that values of the faint-end around $\alpha<-2.3$,  similar to the values reported by \citet[][]{Whitler25}, would lead to an overestimation of the number of sources observed in our field. 
For example, at redshift $z = 10$, using the DPL parameters from previously mentioned paper with $\alpha = -2.6$, the model predicts approximately $n = 4 \pm 1, 18 \pm 4, 65 \pm 10$  and $51\pm8$ sources across the absolute magnitude range of $-18.5$ to $-15.5$ mag. Similarly, at $z = 14$, a DPL fit with $\alpha=-2.42$, also from the same study, would predict $n = 4 \pm 1, 16 \pm 3, 22 \pm 4$, and $20 \pm 4$ sources across the same magnitude range. All of these predictions overestimate our detected number counts.

\subsection{Implications for theoretical models}
\label{sec:implications}

Probing the fainter part of the UV luminosity function at $z>9$ offers, for the first time, critical tests for galaxy formation models that have not been previously tested with such a constraint. For each of the UVLFs shown in Figure \ref{fig:UVLF}, we compare to both pre-\jwst\ and the most recent \jwst-informed theoretical predictions. Since most pre-\jwst\ models underestimate the bright end of the luminosity function \citep[e.g.,][]{castellano23,chemerynska24,donnan24,austin23}, and are based on the assumption of a Schechter function form for the UVLF, they are likely to overestimate the slope at the faint-end. This is the case, for instance, with the model of \citet[][]{mason15}, which shows a much steeper faint-end slope than our currently derived constraints. The model closely tracks the evolution of the dark matter halo mass function, and assuming a redshift-independent star formation efficiency and mass-to-light ratio. It is calibrated using observations at a specific redshift of around $z\approx5$ and predicts a progressive steepening of the faint-end slope, reaching $\alpha=-3.5$ by $z=16$. Consequently, while the model aligns well with the observed LF at $z\sim10$, it significantly overestimates the abundance of faint galaxies and underestimates the number of bright galaxies at higher redshifts. Based on our results, the steepening of the faint-end slope observed at $z < 9$ appears to slow down significantly at $z > 9$. Our finding highlights the overall discrepancy between pre-JWST predictions—largely based on extrapolations of the redshift evolution of the faint-end slope—and \jwst\ observations. 

To reconcile the models with the lack of evolution in the bright end of the LF at $z > 9$, \citet[][]{ferrara23} proposed that the unusual brightness of these sources could be explained by a lack of dust attenuation. This model (dotted line in Figure \ref{fig:UVLF_all}) aligns well with the observed LF at $z \sim 11-12$. However, the model exhibits a significantly stronger redshift evolution compared to our GLIMPSE observations. This results in an overestimate of the galaxy number density at $z=9$ and a substantial underestimate of the LF at $z>12$ especially at the bright end, where dust attenuation is expected to have a more pronounced impact.

\begin{table}
    \centering
    \caption[c]{Binned ultraviolet luminosity function at 9 $< z < 15.$ 
    Column M$_{\rm UV}$ shows the bin in absolute UV magnitude. Column N$_{\rm obj}$  indicates the number of objects within each magnitude bin, and Column $\log (\phi)$ presents the number density of these objects, along with associated uncertainties in the number density.}
        \small{
            \begin{tabular}{lcc}
	       \hline
              \hline
	    M$_{\rm UV}$ & N$_{\rm obj}$  & $\log (\phi)$\\
                 &            &  [mag$^{-1}$Mpc$^{-3}$]\\
	       \hline
              \hline
                &$8.5<z<9.5$&\\
              $-18.5$ & 2  & $-3.48^{+0.34}_{-0.60}$\\
              $-17.5$ & 12 & $-2.68^{+0.21}_{-0.25}$\\
              $-16.5$ & 28 & $-2.26^{+0.18}_{-0.19}$\\
              $-15.5$ & 6  & $-2.34^{+0.24}_{-0.31}$\\
              $-14.5$ & 2  & $-1.99^{+0.32}_{-0.59}$\\
              $-13.5$ & 1  & $-1.41^{+0.39}_{-1.00}$\\
              $-12.5$ & 2  & $-0.03^{+0.31}_{-0.58}$\\
	   \hline
                &$9.5<z<10.5$&\\
              $-18.5$ & 1 & $-3.74^{+0.40}_{-1.01}$\\
              $-17.5$ & 5 & $-3.01^{+0.26}_{-0.35}$\\
              $-16.5$ & 5 & $-2.93^{+0.25}_{-0.34}$\\
              $-15.5$ & 4 & $-2.29^{+0.27}_{-0.38}$\\
              $-13.5$ & 1 & $-1.22^{+0.38}_{-1.00}$\\
              $-12.5$ & 1 & $-0.12^{+0.38}_{-1.00}$\\
	   \hline
                &$10.5<z<11.5$&\\
              $-17.5$ & 4 & $-3.06^{+0.29}_{-0.40}$\\
              $-16.5$ & 6 & $-2.58^{+0.25}_{-0.32}$\\
              $-15.5$ & 2 & $-1.93^{+0.24}_{-0.29}$\\
	   \hline
                &$11.5<z<15$&\\
              $-18.5$ & 1 & $-4.16^{+0.44}_{-1.02}$\\
              $-17.5$ & 4 & $-3.53^{+0.31}_{-0.41}$\\
              $-16.5$ & 5 & $-3.00^{+0.29}_{-0.37}$\\
              $-15.5$ & 1 & $-3.01^{+0.40}_{-1.01}$\\
	   \hline
        \end{tabular}}	
\label{tab: UV_LF}
\end{table}

\begin{table*}
    \centering
        \caption[c]{The best-fitting parameters are derived for DPL fits at each redshift bin. The star formation rate density ($\rho_{\rm {SFR}}$) is calculated by integrating down to $M_{\rm{UV}} = -16$ mag.
        }
    \begin{tabular}{lcccccc}
    \hline
    \hline
          $z$ & $\log (\phi)$  & $M^{*}$  & $\alpha$ & $\beta$ & $\log (\rho_{\rm UV})$ & $\log (\rho_{\rm SFR})$  \\
         & [mag$^{-1}$Mpc$^{-3}$]  & [mag]  & & & [ergs s$^{-1}$ Hz$^{-1}$ Mpc$^{-3}$ ] & [M$_{\odot}$yr$^{-1}$Mpc$^{-3}$]\\
    \hline
    \hline
         8.5-9.5   & $-4.01 _{-0.04}^{+0.04}$ & $-20.30_{-0.20}^{+0.19}$ & $-2.01 _{-0.20}^{+0.20}$ & $-4.30 _{-0.21}^{+0.21}$ & $25.37 \pm 0.44$ & $-2.57 \pm 0.44$  \\
         9.5-10.5  & $-4.19 _{-0.06}^{+0.06}$ & $-20.20_{-0.18}^{+0.19}$ & $-2.07 _{-0.20}^{+0.19}$ & $-4.31 _{-0.19}^{+0.19}$ & $25.20 \pm 0.44$ & $-2.74 \pm 0.44$  \\
         10.5-11.5  & $-4.53 _{-0.01}^{+0.01}$ & $-20.80_{-0.18}^{+0.19}$ & $-2.10 _{-0.19}^{+0.19}$ & $-4.35 _{-0.19}^{+0.19}$ & $25.19 \pm 0.44$ & $-2.75 \pm 0.44$  \\
         11.5-15  & $-4.70 _{-0.21}^{+0.20}$ & $-20.60_{-0.18}^{+0.18}$ & $-2.10 _{-0.19}^{+0.19}$ & $-4.00 _{-0.18}^{+0.20}$ & $24.92 \pm 0.43$ & $-3.02 \pm 0.43$  \\
    \hline
    \end{tabular}
    \label{tab:param}
\end{table*}

Similarly, stochastic star formation histories have been suggested as a possible explanation for the high number density of bright galaxies. Indeed, initial \jwst\ observations have provided empirical evidence supporting bursty star formation histories in early galaxies \citep{strait23,looser23,dressler24,ciesla24,dome24,endsley24}. Significant variations in the star formation rate can cause lower-mass galaxies to experience spikes in UV emission, pushing them to the bright end of the luminosity function. Different implementations of this scenario have been explored, yielding contrasting results. 
\citet{munoz23} and \citet[][]{shen23} explored how adding a scatter to UV magnitudes of individual galaxies affects the UVLF, finding that in the context of $\Lambda$CDM, a scatter of $\sigma_{UV}\gtrsim 1.8$ magnitudes is needed to reproduce the UV excess observed with the JWST at $z\gtrsim 9$. Simulation works exploring this scatter also find highly differing results. \citet[][]{sun2023} find that scatter arising from bursty SFHs in their FIRE simulations is essential and sufficient to reproduce the observed UV excess, while \citet[][]{pallottini23} find a scatter of only $\sigma_{UV} \sim 0.6$ in their {\sc SERRA} simulations. In general, simulations are heavily dependent on the adopted resolution as well as assumed sub-grid physics for star formation and feedback, thus their predictions need much more investigation before convergence can be reached on the strength of the scatter. 
In Figure \ref{fig:UVLF}, we observe that the LF from the stochastically star-forming simulations of \citet[][]{sun2023} is slightly lower than the observations across $z=9-15$, possibly due to a different LF normalization. However, the faint-end shape more closely aligns with observations. These effects may stem from their approach, which combines zoom-in simulations with an extrapolation of the LF from a limited number of dark matter halos to a cosmological volume. In the same plot, we also compare our results with the {\sc SPHINX} cosmological simulations \citep[][]{rosdahl22}, which tend to over-predict the number density of faint galaxies, despite having star-formation and feedback models that yield very stochastic star formation histories (see \citet[][]{Katz23}).

\section{Cosmic UV Luminosity and SFR Density Across redshifts 9 to 15}
\label{sec:SFRD}

\begin{figure*}
    \centering
    \includegraphics[width=\linewidth]{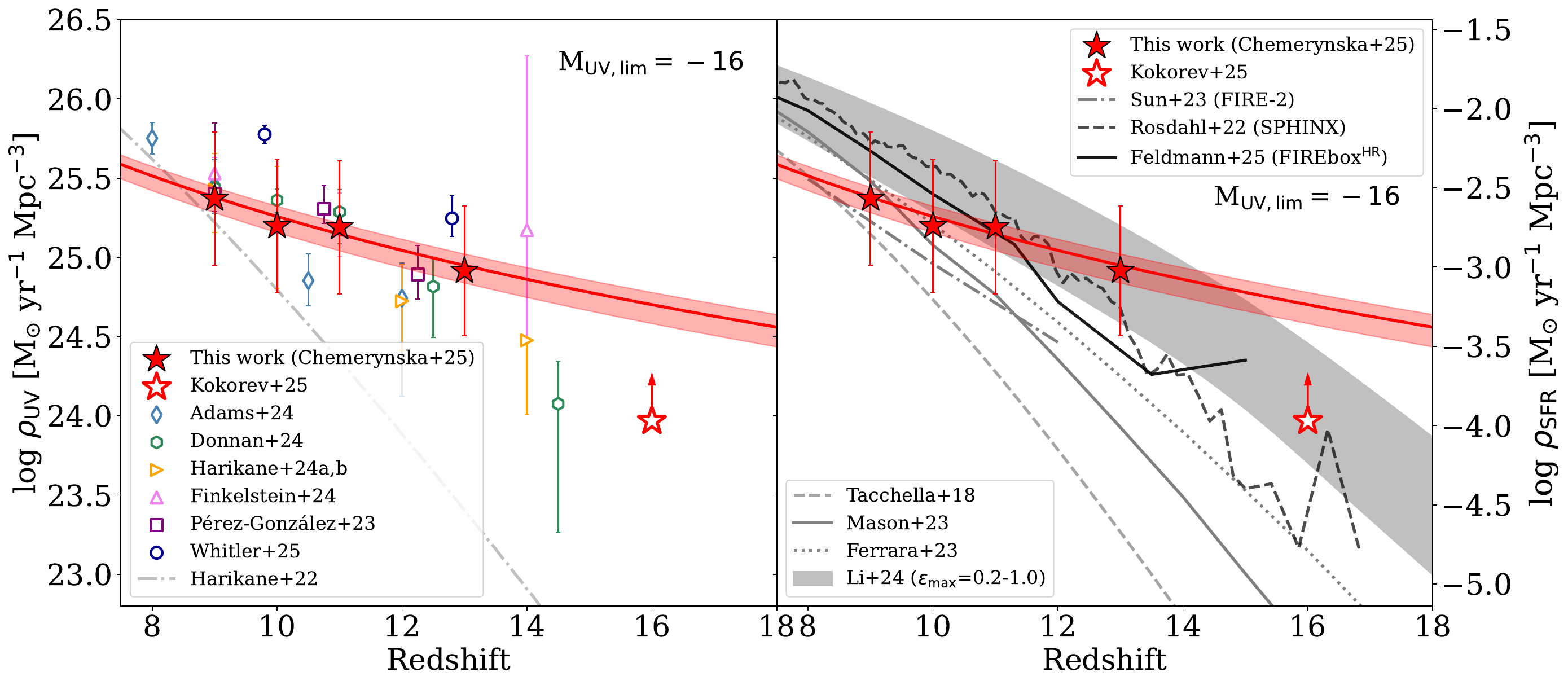}
    \caption{The UV luminosity density (\rhouv\ ) and cosmic star formation rate density (SFRD, \rhosfr\ ) evolution across $z=$9 -- 13. Our \rhouv\ results, hence \rhosfr\ , are shown with red stars and derived by integrating each LF down to \muv$=-16$ mag, including lower limit at $z=16$ from \citet{kokorev25}. We then convert the UV luminosity density \rhouv\ to the SFR density assuming a \citet[][]{salpeter55} IMF with conversion factor $K_{\rm UV}$ = 1.15$\times$10$^{-28}$ M$_{\odot}$ yr$^{-1}$/erg s$^{-1}$ Hz$^{-1}$ (see Table \ref{tab:param}). Our best-fit relation to the redshift-evolution of \rhouv\ is represented by the red line, along with the associated uncertainties represented with the light red region. \textbf{Left:} In addition, we plot literature results with open symbols, including \citet[][]{adams24,donnan24,Harikane24a,Harikane2024,finkelstein23,Perez-Gonzalez23, Whitler25} and fit to the observed SFRD from \citet[][grey dash-dotted line]{harikane22}. \textbf{Right:} We also compare \rhouv\ to the theoretical predictions of \citet[][grey dashed line]{tacchella18}, 
    \citet[][dark grey solid line]{mason23}, \citet[][grey dotted line]{ferrara23}, \citet[][grey shaded region]{li23}, \citet[][dark grey dash-dotted line]{sun2023}, \citet[][black dashed line]{rosdahl22} and \citet[][black solid line]{Feldmann25}, for which we adopt the same integration limit.   
    }
    \label{fig:SFRD}
\end{figure*}

\subsection{The redshift-evolution of the UV luminosity density}

We can derive the total UV luminosity density \rhouv\ of galaxies at all redshifts by integrating the UVLF down to a limiting luminosity set by the galaxy formation physics, or more realistically by observational limits. The limited depth of current \jwst\ programs means that they fail to accurately account for the contribution of lower-mass halos to the total UV luminosity density, which is typically integrated down to \muv $=-17$ mag \citep[][]{harikane22}. The GLIMPSE observations provide a better picture of star formation density and efficiency at early epochs, by accounting for galaxies as faint as \muv $=-13$ mag. Using the best-fit DPL function at each redshift to our UVLF, we integrate the luminosity functions down to a magnitude limit of \muv $ = -16$ mag. This limit has been adopted to maintain a consistent comparison with previously published results, we also integrate them to the same limit \citep{adams24,donnan24,Harikane2024,finkelstein23}. The uncertainties were determined using a statistical method. We ran the MCMC algorithm 100 times, with each run perturbing the data points within their error bars and accounting for redshift bin uncertainties. For each set of parameters, we integrated the UVLF fit to calculate the corresponding value of \rhouv. The final value for \rhouv is the average of these 100 calculations, and its uncertainty is given by the standard deviation of this distribution.

Figure \ref{fig:SFRD} shows the evolving \rhouv\ across the redshift range $z=$9 -- 13 and how it compares to a range of results from the literature and the predictions of galaxy formation models. We provide the derived UV luminosity density in Table \ref{tab:param}. Our results are consistent with most of the literature between $z=9$ and $z=12$ \citep[][]{donnan24,harikane21,finkelstein23}. At higher redshifts, the derived \rhouv\ exceeds most determinations, with the exception of \citet[][]{finkelstein23}. \citet[][]{Whitler25} results on the UV luminosity density are higher at all redshifts. \citet[][]{Perez-Gonzalez23} also find a shallower evolution of the SFRD at $z=8-13$, leading to a factor of 10 discrepancy with most models. We obtain a best-fit relation to the redshift-evolution of the \rhouv, $\rho_{\rm UV} = (2.09^{+0.05}_{-0.12})\times 10^{28}~(1+z)^{-2.94^{+0.06}_{-0.10}}$, that is significantly shallower than most theoretical predictions. 

Thanks to unprecedented constraints on the faint-end of the UVLF, we can, for the first time, integrate the UV luminosity density to much fainter magnitudes to account for the contribution of fainter galaxies. A census of star-forming galaxies down to \muv  $=-13$ mag (Figure \ref{fig:SFRD_13}) shows that the faint galaxy population, with magnitudes between \muv $ = -16$ and $-13$, contributes approximately 50\% of the total UV luminosity density of the Universe across nearly all redshifts beyond $z=9$.  

\subsection{The evolution of cosmic SFR density}

The UV luminosity density can in turn be converted to the star formation rate density (SFRD), noted $\rho_{\star}$ or $\rho_{\text{SFR}}$, assuming a conversion factor between the UV luminosity and the star formation rate based on stellar population models for a given initial mass function (IMF). In the following, we adopt an SFR conversion factor of $K_{\rm UV}$ = 1.15$\times$10$^{-28}$ M$_{\odot}$ yr$^{-1}$/erg s$^{-1}$ Hz$^{-1}$ \citep{Madau14}, assuming a \citet[][]{salpeter55} IMF. This conversion assumes a constant star formation history, which is a clear simplification given the bursty nature of star formation in high-redshift galaxies, especially in low-mass systems \citep[e.g.][]{Tacchella20,asada24,chemerynska24b,Marszewski24,endsley24, looser25}. Rapid variations in star formation can lead to highly fluctuating UV luminosities that may not accurately reflect the averaged star formation rate (SFR) history of these galaxies. Keeping this caveat in mind, the approach enables comparison with recent \jwst\ measurements and theoretical predictions (right axis of Figure \ref{fig:SFRD}), using a consistent conversion factor. 
 
\begin{figure}
    \centering    \includegraphics[width=\linewidth]{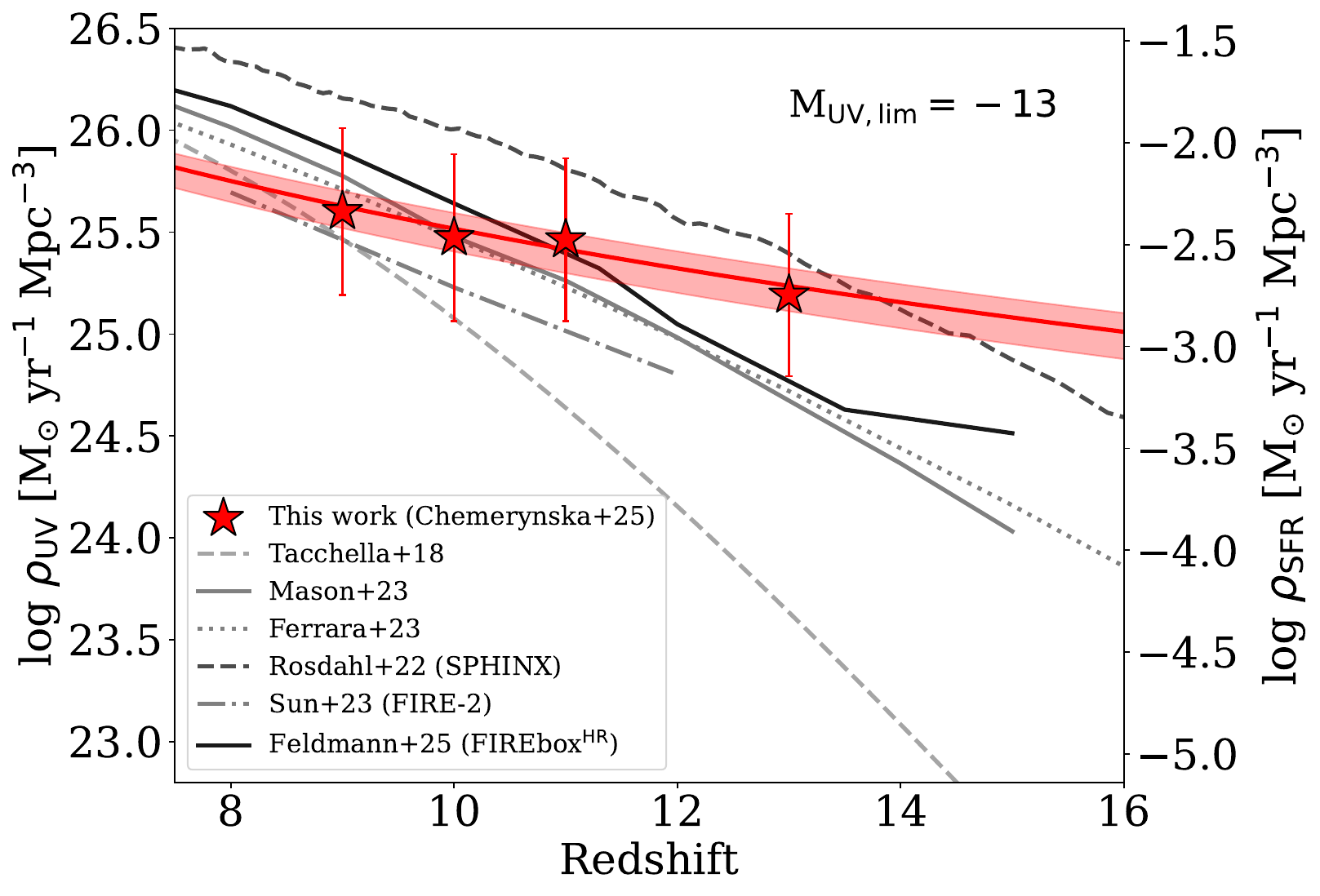}
    \caption{The UV luminosity density and the star formation rate density (SFRD) at $z=9-13$. Our results are shown with the red stars and are based on the integration of the luminosity function at each redshift range, down to an extremely faint-magnitude of \muv $ =-13$ mag. For comparison, we also extend the theoretical models and simulations down to the same integration limit: \citet[][grey dashed line]{tacchella18}
    , \citet[][dark grey solid line]{mason23}, \citet[][grey dotted line]{ferrara23}, \citet[][grey dash-dotted line]{sun2023} and \citet[][black dashed line]{rosdahl22}.
    }
    \label{fig:SFRD_13}
\end{figure}

A direct comparison of the redshift-evolution of the SFRD with the growth of underlying dark matter halo distribution holds key information about the baryon conversion efficiency, i.e., how effectively gas is converted into stars within halos of different masses and at different epochs \citep[][]{tacchella18}. The derived slope of the SFRD-redshift evolution, $\rho_{\rm SFR} = 2.40^{+0.06}_{-0.14} ~ (1+z)^{-2.94^{+0.06}_{-0.10}}$, is significantly shallower than most models, indicating a more gradual increase in the earliest phases of galaxy formation. While recent models match the observed SFRD at $z=9$, they diverge at higher redshifts because of a much steeper slope in the SFRD$-z$ relation, significantly under-predicting the SFRD. Our derived star formation rate density (SFRD) at redshift $z = 13$ is approximately 1 to 2 orders of magnitude higher than the predictions made by \citet[][]{mason23} and \citet[][]{tacchella18}, respectively. These results suggest a higher star formation efficiency than what was predicted at $z > 11$, according to both pre- and post-\textit{JWST} models. However, alternative physical scenarios could also explain these findings. The shallower slope of the SFRD relative to the halo mass function may be attributed to an evolving stellar mass-to-light ratio, which results from a younger stellar population at higher redshifts. 
Figure \ref{fig:SFRD_13} illustrates our results as a function of redshift compared to various model predictions, similar to Figure \ref{fig:SFRD}, which have also been integrated down to the same magnitude limit of \muv$=-13$ mag. Despite clear mismatches with the observed luminosity functions at $z>9$, the models by \citet[][]{mason23}, {\sc Sphinx$^{20}$} \citep[][]{rosdahl22} and FIREbox$^{\mathrm{HR}}$ \citep[][]{Feldmann25} predict values close to our determinations. However, the redshift evolution in these models remains too steep. 
In the following, we provide a detailed comparison between this observational determination and individual models and numerical simulations.

\subsection{Comparison to theoretical models}
\label{sec:comparison}

\begin{table}
    \centering
        \caption{The UV luminosity density and cosmic SFR density integrated down to \muv = $-13$ mag.}
    \begin{tabular}{lcc}
    \hline
    \hline
          $z$ &  $\log (\rho_{\rm UV})$ & $\log (\rho_{\rm SFR})$  \\
            &  [ergs s$^{-1}$ Hz$^{-1}$ Mpc$^{-3}$ ] & [M$_{\odot}$yr$^{-1}$Mpc$^{-3}$]\\
    \hline
    \hline
         8.5-9.5 & $25.60 \pm 0.44$ & $-2.34 \pm 0.44$ \\
        9.5-10.5 & $25.47 \pm 0.44$ & $-2.47 \pm 0.44$ \\
        10.5-11.5 & $25.46 \pm 0.44$ & $-2.48 \pm 0.44$ \\
        11.5-15 & $25.19 \pm 0.43$ & $-2.75 \pm 0.43$ \\
    \hline
    \end{tabular}
    \label{tab:param_13}
\end{table}

Overall, galaxy formation models require an observational anchor for their calibration. Numerical simulations, for instance, rely on phenomenological recipes to represent physical processes occurring on scales smaller than the numerical resolution. This includes crucial components such as star formation and stellar feedback. This ``subgrid'' physics depends on parameters that are adjusted and calibrated against observations to ensure realistic predictions. Several models assume a constant star formation efficiency \citep[][]{mason15,tacchella18,harikane22} as a function of cosmic time, although a dependence with halo mass can be introduced \citep{mason15}. In this framework, the cosmic SFR naturally tracks the growth of dark matter halo mass, leading to a rapid decline in the SFRD at $z>9$. However, our results suggest either an increase in star formation efficiency during the earliest stages of galaxy formation, a modified mass-to-light ratio in these galaxies driven by an evolving IMF \citep[][]{steinhardt23,harikane23}, a high dispersion in the $M_{\star}$-\muv\ relation due to highly stochastic star formation histories \citep[][]{mason23,munoz23,shen23}, feedback-free starbusts~\citep[][]{dekel23,li23}, or dust-free attenuation~\citep[][]{ferrara23}.
Let us explore each of these in turn.

\begin{figure*}
    \centering
    \includegraphics[width=\linewidth]{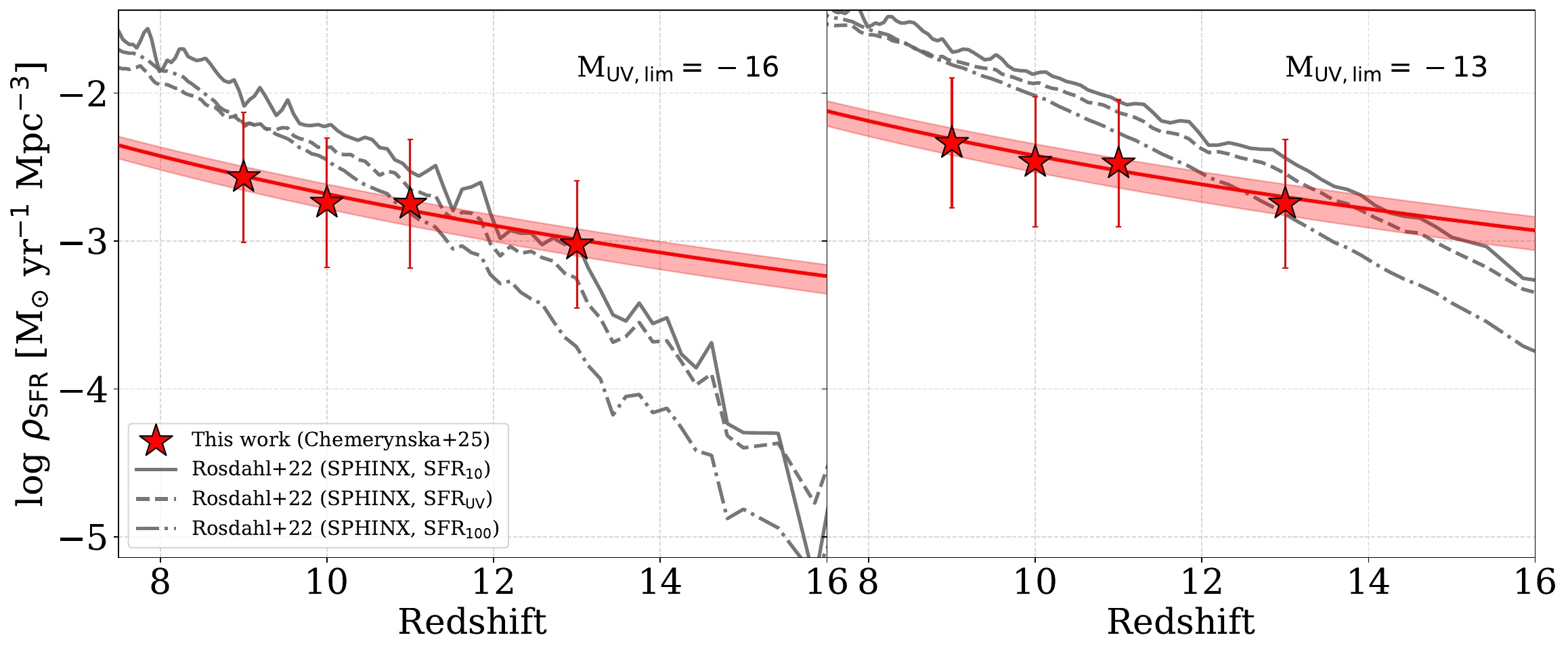}
    \caption{The star formation rate density (SFRD) evolution in the SPHINX simulation \citep{rosdahl22}. We compare our results of \rhosfr\ to different SFRDs derived from SPHINX simulations integrated down to \muv $ =-16$ mag on the left and \muv $=-13$ mag on the right: averaged on 10 Myr timescale (solid gray line), converted from the UV luminosity density (dashed gray line), and averaged on 100 Myr (gray dash-dotted line). 
    }
    \label{fig:SFRD_sphinx}
\end{figure*}

\subsubsection{Star Formation History}
\label{sec:sfh}

Bursty SFHs have recently been explored as a potential explanation for the enhanced UV luminosity density observed at $z>9$ by \jwst. In particular, using zoom-in simulations from {\sc FIRE-2}, \citet[][]{sun2023} demonstrated that a realistic implementation of stellar feedback naturally results in highly bursty star formation. As shown in Figure \ref{fig:UVLF}, such rapid variations in star formation can reproduce the overabundance of UV-bright galaxies (\muv$<-20$) at $z=9$. However, this approach struggles to match both the fainter end of the LF and its redshift evolution. Consequently, this SFH-driven model shows poor agreement with our measurements when comparing the integrated UV luminosity density as a function of redshift (Figure \ref{fig:SFRD} and \ref{fig:SFRD_13}).     

Additionally, the FIREbox$^{\rm{HR}}$ simulation \citep{Feldmann25} demonstrates good agreement with observational data. Similar to {\sc FIRE-2}, it reveals that relatively high star formation activity is common in the early Universe. FIREbox$^{\rm{HR}}$ resolves a significant number of faint, low-mass galaxies. The simulation suggests a non-evolving/weak mass-dependent SFE, which accounts for the increased UV luminosity densities observed at redshifts of $z\sim6-14$. Their results imply that a nearly constant SFE for intermediate-mass halos ($M_{\rm{halo}}\sim10^9-10^{11}$ \msol) aligns well with observations, without requiring highly bursty SFHs. This leads to a larger contribution from lower-mass halos at higher redshifts, contributing to the evolution of cosmic UV luminosity density. This contrasts with that of {\sc FIRE-2}, which attributes the UV-bright galaxy population to rapid bursts of star formation induced by stellar feedback.

Given the uncertainties associated with the conversion from the UV luminosity to the SFR density, we investigate the effects of star formation timescales on the evolution of the SFRD in the SPHINX hydrodynamical simulations and how they compare to our derived constraints. Figure \ref{fig:SFRD_sphinx} illustrates the SFRD estimated from three distinct indicators in the {\sc Sphinx$^{20}$} simulations \citep[][]{rosdahl22}. The solid line represents the SFRD averaged over 10 Myr, providing a more realistic estimate for bursty SFHs. The dot-dashed line represents the SFRD averaged over the last 100 Myr, which is significantly lower than the SFRD${10}$ line. This highlights the bursty nature of these galaxies during early epochs, resulting in much brighter luminosities compared to values averaged over longer timescales. Additionally, we present the SFRD derived from the UV luminosity in SPHINX, assuming the same conversion factor as used in observations. The dashed curve aligns more closely with SFRD${100}$ at $z=9$, but it exhibits a distinct slope as a function of redshift. This divergence primarily arises from the reduction in dust attenuation with increasing redshift.

\subsubsection{Feedback-free bursts} Among other scenarios proposed in light of recent \jwst\ results, feedback-free starbursts \citep[][]{dekel23,li23} predict an enhanced star formation efficiency, which can explain the observed overabundance of UV-bright galaxies at $z>8$. The model defines a density threshold, $n\sim 3\times 10^{3}$ cm$^{-3}$, above which the cooling time is shorter than the free-fall time, which leads to an enhancement of the star formation efficiency by a factor $\epsilon_{\rm max}=0.2$ to 1. The slow SFRD evolution inferred from our observations means that the SFRD derived at $z=9-10$ is consistent with an efficiency of $\epsilon_{\rm max}=0.2$, while the SFRD derived at $z>12$ requires an efficiency close to $\epsilon_{\rm max} = 1$ (gray-shaded region in figure \ref{fig:SFRD}). In other words, a large increase in the maximum star-formation efficiency with redshift is required for the FFB model to match observations.

\subsubsection{Attenuation-free galaxies} Another scenario proposed in \citet[][]{ferrara23} relies on outflows that become strong enough above a specific SFR threshold to clear the dust/gas content in galaxies. In this model, the increased un-obscured UV luminosity compensates the strong decline in the DM halo mass function, which could explain the overabundance of UV-bright galaxies at $z>9$. Our galaxies show UV continuum slopes around $\beta \sim -2.3$ to $\beta \sim -2.5$ on average across $z=9-15$ (Jecmen et al. in prep.). These rather blue but not extreme slopes are consistent with low-dust attenuation. In Figure \ref{fig:SFRD} the model captures the SFRD$-z$ evolution of most results up to $z\sim 12$ but falls up to 7 times below our constraints at higher redshifts.

\subsection{Caveats} 

While our results clearly show enhanced star-formation activity at $z>9$ compared to theoretical models and simulations, they are based on a photometrically selected sample of galaxies. Consequently, uncertainties remain regarding the true redshifts of these sources, as the possibility of low-redshift contaminants, such as dusty star-forming galaxies, cannot be entirely ruled out \citep[][]{zavala23,naidu22b,arrabal23}. While spectroscopic confirmation is required, the faint population reaches observed magnitudes beyond $m_{\rm AB} = 30, \mathrm{mag}$ in F277W filter, making exposure times prohibitive even for \jwst\ NIRSpec. An observing strategy that samples this population across different magnitudes could help calibrate the photometric redshifts and, importantly, quantify the contamination rate. Furthermore, a better understanding of the fainter dusty star-forming galaxies at $z=4-5$ will also improve the galaxy templates used in photometric redshift estimates \citep{barrufet23,Weibel24}.  These galaxies can mimic the broad-band colors of high-redshift galaxies and be misidentified as high-z candidates \citep[][]{naidu22, haro23}. 

While the UV luminosity density is a direct measurement, the cosmic star formation rate density (SFRD) is computed by assuming standard calibrations based on a \cite{salpeter55} IMF and a constant star formation history. As discussed in Sect. \ref{sec:sfh}, deviations from these assumptions are expected for early and low-mass galaxies. In the case of bursty star formation, as expected in these galaxies, the UV luminosity may not reliably represent the average SFR over that timescale. This is particularly true for UV-selected galaxies, which are often identified during a recent burst of star formation, typically around 10 Myr, as suggested by recent \jwst\ findings \citep[e.g.,][]{looser23}. These bursts can contribute to the global enhancement of \rhouv. Additionally, since lower-mass galaxies are more prone to experiencing bursty SFHs \citep[e.g.][]{atek22}, such effects can also influence the shape of the UV luminosity function.

Furthermore, variations in the IMF (e.g., a top-heavy IMF) alter the fraction of UV-emitting massive stars, impacting the UV-to-SFR conversion factor \citep[][]{chon22, steinhardt22, yung24, Cueto24,Hutter24}. Additionally, this calibration primarily considers single-star evolution, whereas binary interactions can extend the lifetimes of UV-bright stars, modifying the UV luminosity-SFR relationship, especially in low-metallicity environments. These interactions can lead to a 0.15–0.2 dex higher conversion coefficient \citep[e.g.,][]{ma16, wilkins19}. Along similar lines, very massive stars (VMS), typically with masses greater than 100 \msol, may also impact the UV luminosity distribution for a given stellar population \citep[][]{crowther10, eldridge22, schaerer24}. Moreover, changing the IMF primarily rescales the mass-to-light ratio ($M/L$), which in turn affects both stellar masses and star SFRs \citep{madau2014}. Therefore, much like the IMF, the star formation history and stellar age are critical factors that directly influence the $M/L$. For instance, \cite{Santini23} has revealed that galaxies at redshifts greater than $z>7$ exhibit a two orders of magnitude range in $M/L$ values for a given luminosity. It is a strong indicator of the broad variety of physical conditions and diverse SFH present in the early galaxies.

All of these measurements are observed quantities and have not been corrected for dust attenuation. Applying such a correction would increase both the UV luminosity and the resulting cosmic SFR density values. However, considering the observed UV continuum slopes (Section \ref{sec:comparison}), the dust content in these galaxies is likely low, making the impact of dust attenuation negligible compared to the aforementioned effects \citep[][]{austin23,Narayanan24, saxena24,cullen23,Topping24b}.

An additional factor that could influence our results is the size-luminosity relation, which directly affects the completeness of our calculations. In Section \ref{sec:completeness}, we utilized the log-normal distribution derived from \citet[][]{yang2022} and \citet[][]{shibuya15}. These studies provide measurements of galaxy size and luminosity at redshifts ($z \sim 4-15$). Accurately accounting for galaxy sizes is crucial at faint magnitudes, as smaller faint objects tend to be better recovered \citep[][]{atek18}. Consequently, the impact of completeness at the faint-end becomes more significant. Moreover, variations in the size-luminosity relation can significantly affect the slope of the faint-end, potentially making it steeper. Non-parametric size measurements by \citet[][]{curtis-lake16} show no evidence of a positive size-luminosity relation at redshifts $z=4-8$. In this case, the impact of completeness corrections at the faint-end would be more pronounced.

One of the main uncertainties arises from potential field-to-field variations. Given the small volume probed by our survey, particularly in high-magnification regions, cosmic variance could play a significant role, meaning that these properties may not be representative of the large-scale galaxy distribution. This could explain the high number density of galaxies inferred at $z\sim17$ \citep[][]{kokorev25}, while previous equivalently deep blank surveys did not identify any sources at these redshifts. 
However, when calculating the UV luminosity function, we have accounted for statistical variations due to cosmic variance, and the best-fit UVLF used to derive the SFRD incorporates these uncertainties. Moreover, the higher-than-expected cosmic SFR density is not observed in a single volume at a given redshift, but rather at nearly all redshifts beyond $z=9$, which contributes to the observed slow redshift evolution.

\section{Summary}
\label{sec:summary}

We use ultra-deep NIRCam imaging of the lensing cluster Abell S1063 to investigate the faintest galaxy population ever observed at $z>9$. These observations are part of the \jwst\ GLIMPSE survey, which has achieved the deepest imaging of the sky to date, offering an unprecedented view into Cosmic Dawn and the formation and evolution of the earliest galaxies. Using the Lyman-break technique combined with photometric redshifts, we identify a robust sample of 105 galaxy candidates spanning the redshift range 
$9<z<15$. This includes 98 galaxies at $z\sim9-11$ and 7 galaxies at $z\sim11-15$. Notably, some of these sources represent the faintest galaxies ever detected at $z>9$, with absolute UV magnitudes as faint as \muv $\approx -13$, corresponding to approximately 0.0003L$^*$, the characteristic luminosity at $z\sim12$ (Table \ref{tab:param}).

Using these samples, we compute the UV luminosity function (UVLF) across the redshift range $z=9$ to $z=$ 13. Leveraging extensive simulations and our cluster mass model, we carefully analyzed all lensing effects directly in the source plane to quantify their impact on the survey volume and the resulting UVLFs. Given the small survey volume probed by the GLIMPSE survey, it is primarily sensitive to galaxies fainter than \muv$=-19$ mag. To account for this, we combine our faint-end data points with results from wide-area surveys in the literature, which probe the bright end of the LF at \muv$< -18.5$ mag, to provide a comprehensive fit to the UVLF across the full magnitude range.

In the next step, we derive the UV luminosity density, \rhouv, at each redshift by integrating the UV luminosity functions down to \muv $=-16$ mag. We then convert this luminosity density into the cosmic star formation rate density (SFRD), using the calibration of \citep[][]{madau2014} and assuming a \citep[][]{salpeter55} initial mass function. Our main results are summarized as follows:
\begin{itemize}
    \item For the first time, we extend the UV luminosity function down to \muv $\sim $ -14 mag at $z>9$, which is on average 3 magnitudes fainter than any previous \jwst\ observations. We observe minimal evolution of the LF with redshift, as indicated by a faint-end slope that varies from $\alpha=-2.01\pm 0.20$ at $z=8.5-9.5$ to $\alpha=-2.10\pm 0.19$ at $z=11.5-15$. This lack of strong evolution is consistent with recent deep \jwst\ observations that probe brighter galaxies \citep[][]{Perez-Gonzalez23}. Our findings also align with recent \jwst\ results showing slow evolution of the bright end of the LF \citep[][]{bouwens23,finkelstein23,adams24,donnan24}. 

    \item Similarly, our best-fit relation for the redshift evolution of the cosmic SFR density, \rhosfr, at $z>9$ follows a slope of $\propto (1+z)^{-2.94}$. This slope is significantly shallower than most theoretical predictions, including the most recent ones informed by early \jwst\ results. As a consequence, this leads to a clear underestimate of the theoretical models of the SFRD at redshifts beyond $z \sim 12$.
    \item For the first time, we integrate the UV luminosity density and cosmic SFRD down to extremely faint luminosities, \muv$= -13$ mag. As before, the evolution of the SFRD as a function of redshift is slower than predicted by most galaxy formation models. Notably, we find that galaxies fainter than \muv $=-16$ contribute to more than 50\% of the total SFRD in the Universe at redshifts $z = 9-15$. 
\end{itemize}
Overall, our findings suggest an enhancement of star formation efficiency at $z>9$. While the exact physical mechanisms remain uncertain, several plausible scenarios are under investigation. For instance, the observed increase in the UV luminosity density could result from bursty star formation rather than a genuine enhancement of star formation efficiency. In this case, key parameters of the star formation histories—such as the dust cycle and UV dispersion—must be quantified to reconcile the faint-end of the luminosity function and the corresponding evolution of the SFRD at $z>9$. Additionally, an evolving initial mass function (IMF) could alter the UV-to-SFR conversion, leading to an overestimation of the inferred SFR. Finally, while cosmic variance uncertainties have been incorporated into our error estimates, field-to-field variations cannot be completely ruled out. Further ultra-deep observations of similarly lensed fields will be critical to confirm this unexpectedly high UV luminosity density in the early Universe.

\section*{Acknowledgments}\

IC and HA acknowledge support from CNES, focused on the JWST mission, and the Programme National Cosmology and Galaxies (PNCG) of CNRS/INSU with INP and IN2P3, co-funded by CEA and CNES. IC acknowledges funding support from the Initiative Physique des Infinis (IPI), a research training program of the Idex SUPER at Sorbonne Université. HA acknowledges support by the French National Research Agency (ANR) under grant ANR-21-CE31-0838. The BGU lensing group acknowledges support by grant No.~2020750 from the United States-Israel Binational Science Foundation (BSF) and grant No.~2109066 from the United States National Science Foundation (NSF), and by the Israel Science Foundation Grant No.~864/23. VK acknowledges support from the University of Texas at Austin Cosmic Frontier Center. AA acknowledges support by the Swedish research council Vetenskapsradet (2021-05559). 
This work has received funding from the Swiss State Secretariat for Education, Research and Innovation (SERI) under contract number MB22.00072, as well as from the Swiss National Science Foundation (SNSF) through project grant 200020\_207349.
This work has made use of the \texttt{CANDIDE} Cluster at the \textit{Institut d'Astrophysique de Paris} (IAP), made possible by grants from the PNCG and the region of Île de France through the program DIM-ACAV+, and the Cosmic Dawn Center and maintained by S. Rouberol. P.N. acknowledges support from the Gordon and Betty Moore Foundation and the John Templeton Foundation that fund the Black Hole Initiative (BHI) at Harvard University where she serves as one of the PIs. This work is based on observations obtained with the NASA/ESA/CSA \textit{JWST} and the NASA/ESA \textit{Hubble Space Telescope} (HST), retrieved from the \texttt{Mikulski Archive for Space Telescopes} (\texttt{MAST}) at the \textit{Space Telescope Science Institute} (STScI). STScI is operated by the Association of Universities for Research in Astronomy, Inc. under NASA contract NAS 5-26555.

\section*{Data Availability}
The data underlying this article are publicly available on the \texttt{Mikulski Archive for Space Telescopes}\footnote{\url{https://archive.stsci.edu/}} (\texttt{MAST}), under program ID 3293.

\appendix
\section{High-redshift candidates}

\begin{figure*}
    \centering
            \includegraphics[width=0.9\textwidth]{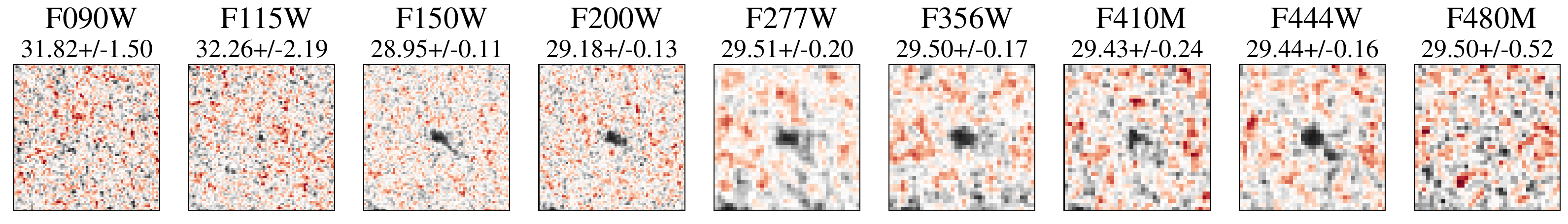}
                \includegraphics[width=0.9\textwidth]{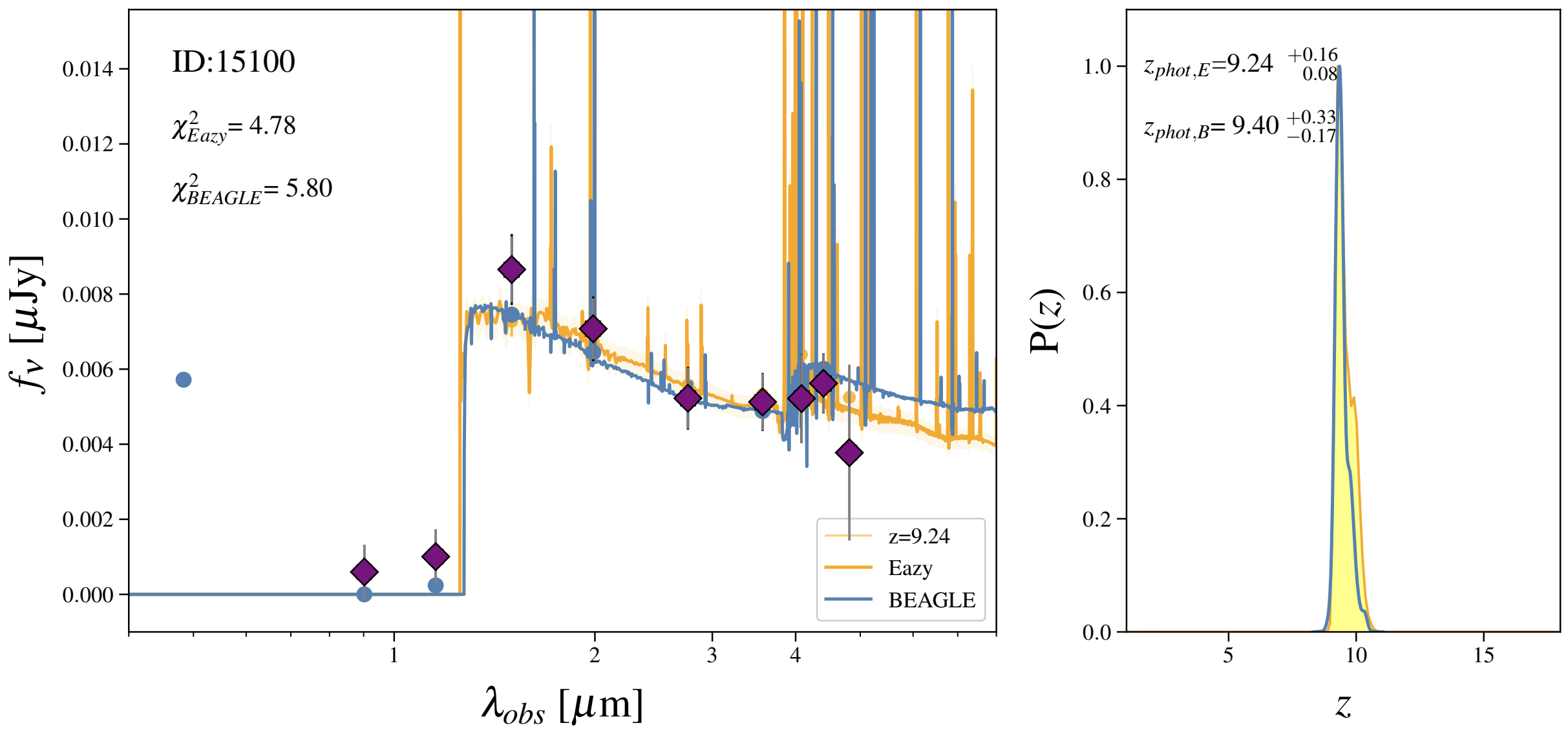} \\
                   \includegraphics[width=0.9\textwidth]{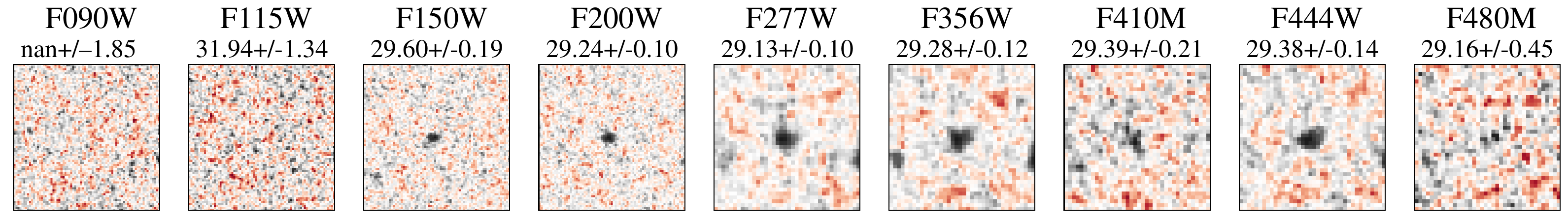} 
                \includegraphics[width=0.9\textwidth]{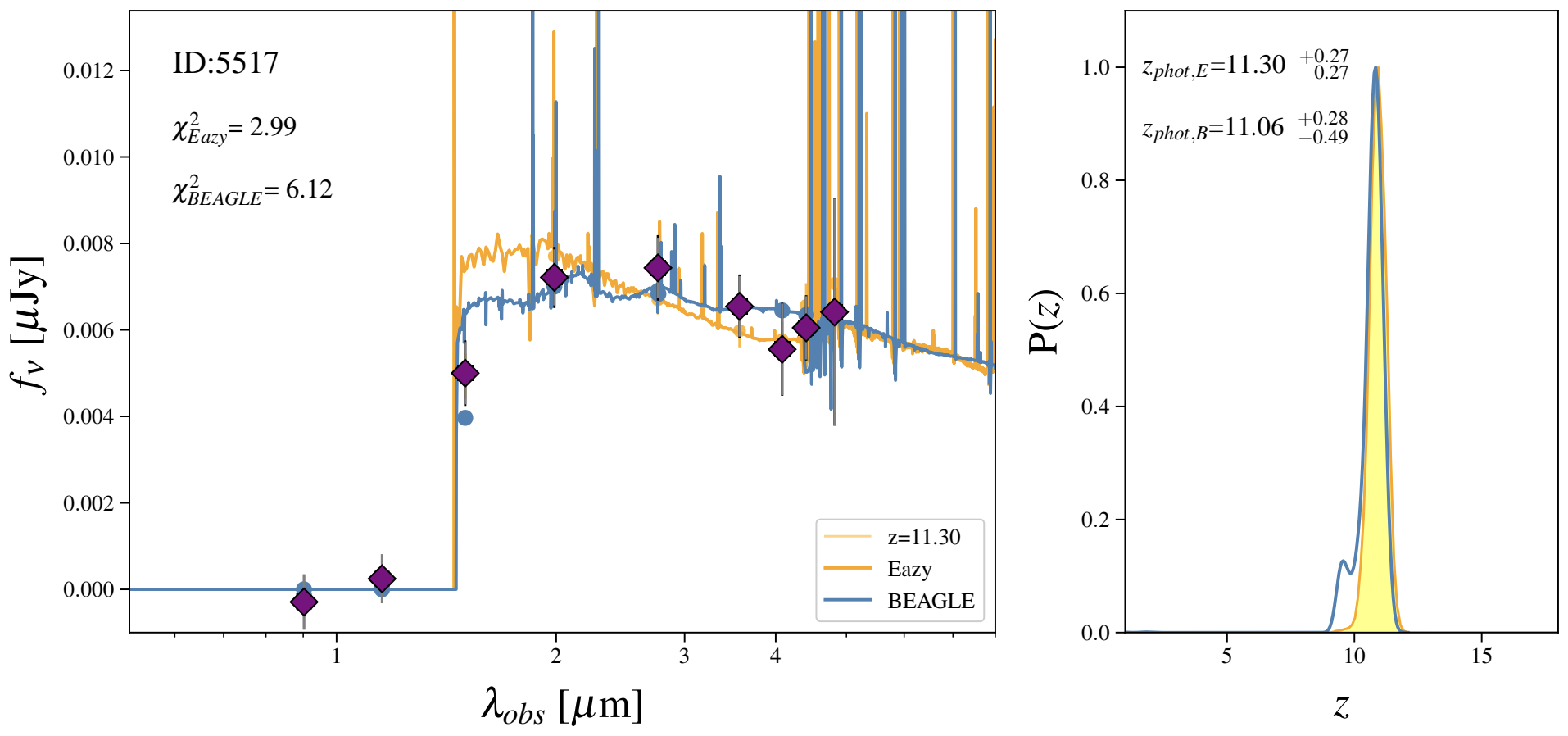}

    \caption{Examples of galaxy candidates at different redshifts, showing NIRCam imaging and best-fit SED solutions. The top row displays cutouts of each candidate across the nine \jwst\ filters. The bottom panel features the best-fit SEDs generated by {\sc Eazy} (orange curve) and {\sc Beagle} (blue curve), along with the object ID and the best-fit $\chi^{2}$ values from both codes. The purple diamonds represent the observed fluxes, while the orange and blue circles show the best-fit model magnitudes from {\sc Eazy} and {\sc Beagle}, respectively. On the right panel, the photometric redshift probability distribution functions (PDFs) from both codes are shown, in addition to the best-fit redshift and associated uncertainties.}
    \label{fig:cutout1}
\end{figure*}

In this section, we provide additional details about the galaxy candidates selected in Section \ref{sec:High-z}. Figure \ref{fig:cutout1} showcases the best-fit SEDs for a representative candidate from each redshift range, derived using two independent codes: {\sc Beagle} and {\sc Eazy}. While at the highest redshifts, the two codes may differ in attributing the flux excess in the reddest band to either a Balmer break or strong emission lines, they consistently converge on the best-fit redshift solution. This agreement is reflected in the narrow posterior probability distribution of the redshift. Additionally, we present imaging cutouts of the candidates in each of the nine NIRCam filters, highlighting the Lyman break and their corresponding magnitudes.

\begin{figure*}
    \centering
                    \includegraphics[width=0.9\textwidth]{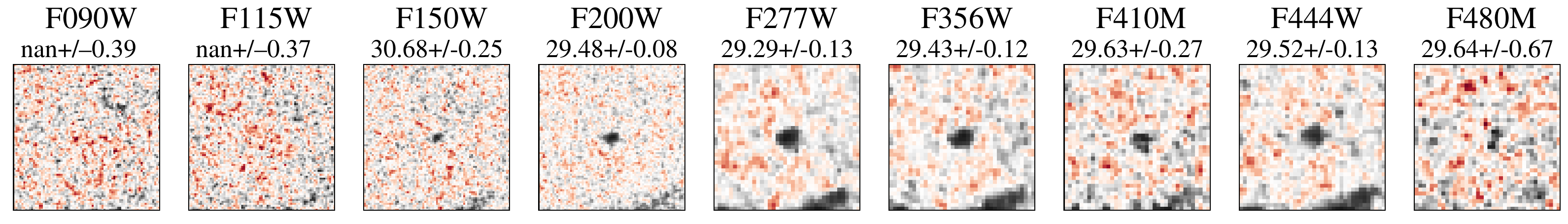}
        \includegraphics[width=0.9\textwidth]{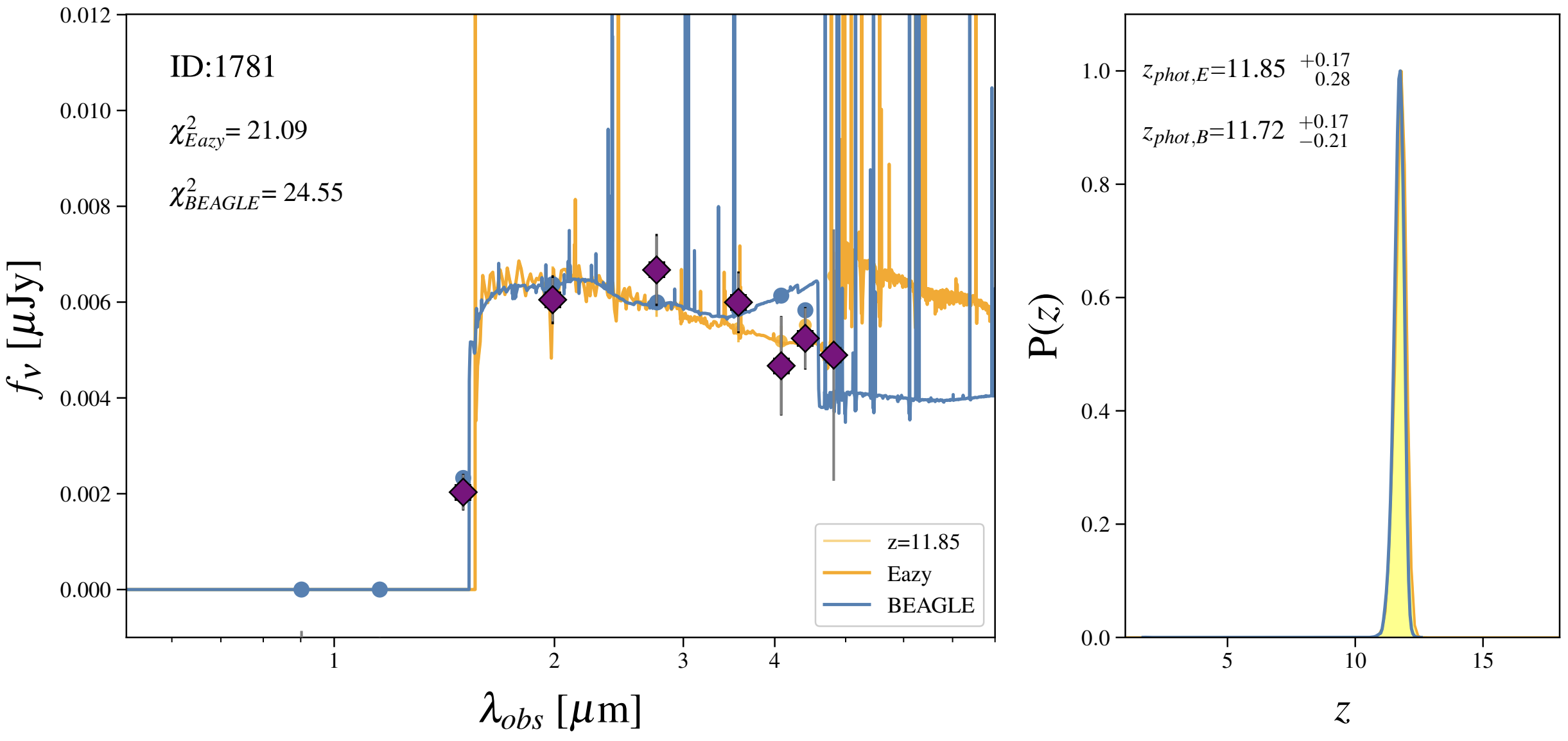} 

    \caption{Same as Figure \ref{fig:cutout1}}
    \label{fig:cutout2}
\end{figure*}



\bibliographystyle{mnras}
\bibliography{reference,refs} 




 \begin{table*}
     \centering
     \caption{Summary of the full sample of the galaxy candidates at $z\sim9-15$ analyzed in this paper.}
     \begin{tabular}{lccccc}
     \hline
     \hline
       ID & RA(J2000)& DEC(J2000)& z$_{\rm phot}$& \muv [mag] & $\mu$\\
      \hline
      \hline 
        5255   &  342.236847  &  -44.560589 & 8.58 $\pm$ 0.61 & -16.40 $\pm$ 0.22 & 1.29 $\pm$ 0.15 \\
        48305  &  342.246552  &  -44.542870 & 8.59 $\pm$ 0.10 & -18.83 $\pm$ 0.04 & 1.37 $\pm$ 0.15 \\
        13557  &  342.250854  &  -44.547131 & 8.63 $\pm$ 0.13 & -16.94 $\pm$ 0.05 & 1.32 $\pm$ 0.15 \\
        56616  &  342.229797  &  -44.542038 & 8.65 $\pm$ 0.08 & -17.95 $\pm$ 0.03 & 1.52 $\pm$ 0.19 \\
        12280  &  342.234436  &  -44.548862 & 8.66 $\pm$ 0.11 & -18.00 $\pm$ 0.04 & 1.39 $\pm$ 0.16 \\
        30848  &  342.205872  &  -44.532963 & 8.73 $\pm$ 0.35 & -15.29 $\pm$ 0.13 & 3.21 $\pm$ 0.48 \\
        3468   &  342.223724  &  -44.564442 & 8.75 $\pm$ 0.12 & -17.63 $\pm$ 0.04 & 1.34 $\pm$ 0.15 \\
        31824  &  342.204987  &  -44.532993 & 8.76 $\pm$ 0.37 & -15.73 $\pm$ 0.14 & 3.37 $\pm$ 0.50 \\
        3492   &  342.227173  &  -44.564388 & 8.77 $\pm$ 3.33 & -16.99 $\pm$ 1.24 & 1.31 $\pm$ 0.15 \\
        52490  &  342.171326  &  -44.516994 & 8.77 $\pm$ 0.64 & -15.92 $\pm$ 0.23 & 2.34 $\pm$ 0.32 \\
        60624  &  342.166107  &  -44.523674 & 8.78 $\pm$ 3.45 & -15.69 $\pm$ 1.36 & 2.94 $\pm$ 0.43 \\
        6208   &  342.213135  &  -44.558884 & 8.82 $\pm$ 0.42 & -16.78 $\pm$ 0.15 & 1.44 $\pm$ 0.18 \\
        1605   &  342.239563  &  -44.569675 & 8.87 $\pm$ 0.41 & -16.86 $\pm$ 0.14 & 1.24 $\pm$ 0.14 \\
        11203  &  342.253265  &  -44.550365 & 8.87 $\pm$ 0.30 & -16.99 $\pm$ 0.10 & 1.29 $\pm$ 0.15 \\
        4683   &  342.238068  &  -44.561813 & 8.88 $\pm$ 0.15 & -17.76 $\pm$ 0.05 & 1.28 $\pm$ 0.14 \\
        7171   &  342.238098  &  -44.557068 & 8.89 $\pm$ 0.56 & -16.66 $\pm$ 0.19 & 1.31 $\pm$ 0.15 \\
        2797   &  342.238190  &  -44.566078 & 8.91 $\pm$ 3.44 & -16.29 $\pm$ 1.26 & 1.26 $\pm$ 0.14 \\
        5833   &  342.236633  &  -44.559528 & 8.95 $\pm$ 0.92 & -16.81 $\pm$ 0.31 & 1.30 $\pm$ 0.15 \\
        40044  &  342.153748  &  -44.537327 & 8.96 $\pm$ 0.10 & -16.44 $\pm$ 0.04 & 3.46 $\pm$ 0.56 \\
        14967  &  342.254303  &  -44.545643 & 8.98 $\pm$ 1.39 & -16.90 $\pm$ 0.48 & 1.31 $\pm$ 0.15 \\
        57891  &  342.206085  &  -44.521862 & 9.00 $\pm$ 0.64 & -12.50 $\pm$ 1.27 & 59.90 $\pm$ 56.75 \\
        69815  &  342.171112  &  -44.530312 & 9.00 $\pm$ 4.38 & -12.05 $\pm$ 2.80 & 87.20 $\pm$ 71.51 \\
        36308  &  342.200836  &  -44.535419 & 9.02 $\pm$ 0.37 & -15.83 $\pm$ 0.13 & 3.52 $\pm$ 0.48 \\
        1177   &  342.240173  &  -44.570992 & 9.05 $\pm$ 0.46 & -17.30 $\pm$ 0.15 & 1.23 $\pm$ 0.14 \\
        72488  &  342.257782  &  -44.551109 & 9.06 $\pm$ 0.59 & -16.69 $\pm$ 0.20 & 1.27 $\pm$ 0.14 \\
        9783   &  342.259094  &  -44.552521 & 9.07 $\pm$ 0.47 & -16.84 $\pm$ 0.16 & 1.26 $\pm$ 0.14 \\
        12322  &  342.248566  &  -44.548744 & 9.07 $\pm$ 0.28 & -17.37 $\pm$ 0.09 & 1.32 $\pm$ 0.15 \\
        37282  &  342.195984  &  -44.535862 & 9.08 $\pm$ 1.20 & -14.63 $\pm$ 0.45 & 4.97 $\pm$ 0.76 \\
        1837   &  342.249146  &  -44.568836 & 9.09 $\pm$ 0.49 & -16.97 $\pm$ 0.16 & 1.22 $\pm$ 0.14 \\
        10705  &  342.257782  &  -44.551113 & 9.09 $\pm$ 0.69 & -16.69 $\pm$ 0.23 & 1.27 $\pm$ 0.14 \\
        56793  &  342.178528  &  -44.520859 & 9.10 $\pm$ 1.29 & -14.79 $\pm$ 0.48 & 4.56 $\pm$ 0.76 \\
        2416   &  342.239655  &  -44.567123 & 9.12 $\pm$ 0.36 & -16.92 $\pm$ 0.12 & 1.25 $\pm$ 0.14 \\
        38745  &  342.198151  &  -44.536610 & 9.12 $\pm$ 0.08 & -16.51 $\pm$ 0.03 & 3.75 $\pm$ 0.57 \\
        37840  &  342.233521  &  -44.536118 & 9.16 $\pm$ 0.93 & -16.29 $\pm$ 0.31 & 1.61 $\pm$ 0.19 \\
        14073  &  342.239227  &  -44.546658 & 9.17 $\pm$ 1.80 & -17.45 $\pm$ 0.61 & 1.38 $\pm$ 0.16 \\
        54451  &  342.198242  &  -44.519318 & 9.19 $\pm$ 0.58 & -13.29 $\pm$ 2.29 & 53.32 $\pm$ 111.3\\
        41319  &  342.242310  &  -44.538002 & 9.22 $\pm$ 0.72 & -16.88 $\pm$ 0.24 & 1.45 $\pm$ 0.18 \\
        12594  &  342.273621  &  -44.548538 & 9.24 $\pm$ 0.66 & -18.34 $\pm$ 0.22 & 1.23 $\pm$ 0.13 \\
        15100  &  342.173767  &  -44.545620 & 9.24 $\pm$ 0.12 & -16.67 $\pm$ 0.05 & 5.48 $\pm$ 0.80 \\
        62329  &  342.164032  &  -44.525223 & 9.26 $\pm$ 0.52 & -16.51 $\pm$ 0.18 & 3.04 $\pm$ 0.44 \\
        1567   &  342.246490  &  -44.569771 & 9.29 $\pm$ 3.67 & -16.61 $\pm$ 1.28 & 1.22 $\pm$ 0.13 \\
        10727  &  342.272583  &  -44.551075 & 9.31 $\pm$ 0.23 & -18.00 $\pm$ 0.07 & 1.23 $\pm$ 0.13 \\
        2360   &  342.242371  &  -44.567291 & 9.33 $\pm$ 1.07 & -16.67 $\pm$ 0.35 & 1.24 $\pm$ 0.14 \\
        41930  &  342.233246  &  -44.538322 & 9.33 $\pm$ 0.37 & -17.10 $\pm$ 0.12 & 1.56 $\pm$ 0.19 \\
        1797   &  342.256927  &  -44.568947 & 9.35 $\pm$ 0.61 & -16.40 $\pm$ 0.20 & 1.21 $\pm$ 0.13 \\
        1675   &  342.252380  &  -44.569401 & 9.37 $\pm$ 0.74 & -16.52 $\pm$ 0.24 & 1.21 $\pm$ 0.13 \\
        73377  &  342.242371  &  -44.567303 & 9.39 $\pm$ 3.85 & -16.72 $\pm$ 1.34 & 1.24 $\pm$ 0.14 \\
        11464  &  342.234955  &  -44.549992 & 9.41 $\pm$ 0.65 & -17.74 $\pm$ 0.21 & 1.38 $\pm$ 0.17 \\
        57073  &  342.161499  &  -44.521320 & 9.41 $\pm$ 0.27 & -16.79 $\pm$ 0.09 & 2.21 $\pm$ 0.30 \\
        70473  &  342.198578  &  -44.535030 & 9.41 $\pm$ 0.41 & -15.92 $\pm$ 0.14 & 4.42 $\pm$ 0.74 \\
        3268   &  342.242767  &  -44.564835 & 9.43 $\pm$ 4.18 & -17.20 $\pm$ 1.46 & 1.25 $\pm$ 0.14 \\
        47627  &  342.235321  &  -44.542278 & 9.43 $\pm$ 0.44 & -16.63 $\pm$ 0.14 & 1.46 $\pm$ 0.17 \\
        5710   &  342.214111  &  -44.559807 & 9.44 $\pm$ 0.28 & -17.35 $\pm$ 0.09 & 1.43 $\pm$ 0.17 \\
        \hline
        63930  &  342.161499  &  -44.525753 & 9.52 $\pm$ 0.03 & -18.27 $\pm$ 0.01 & 2.82 $\pm$ 0.43 \\
        13276  &  342.181305  &  -44.547470 & 9.56 $\pm$ 0.51 & -15.55 $\pm$ 0.17 & 3.00 $\pm$ 0.41 \\
        55592  &  342.182129  &  -44.520252 & 9.67 $\pm$ 0.42 & -16.36 $\pm$ 0.15 & 7.14 $\pm$ 1.28 \\
        48785  &  342.160278  &  -44.541546 & 9.88 $\pm$ 0.38 & -15.26 $\pm$ 0.13 & 7.72 $\pm$ 1.52 \\
        1791   &  342.236115  &  -44.568989 & 9.89 $\pm$ 0.38 & -17.14 $\pm$ 0.12 & 1.25 $\pm$ 0.14 \\
        4357   &  342.231903  &  -44.562424 & 9.92 $\pm$ 0.50 & -16.50 $\pm$ 0.15 & 1.30 $\pm$ 0.15 \\
        48074  &  342.195038  &  -44.542713 & 9.99 $\pm$ 3.95 & -15.56 $\pm$ 1.34 & 2.63 $\pm$ 0.36 \\
        10544  &  342.257294  &  -44.551338 & 10.07 $\pm$ 4.35 & -16.65 $\pm$ 1.41 & 1.27 $\pm$ 0.14 \\
        22732  &  342.234650  &  -44.530128 & 10.10 $\pm$ 0.53 & -17.00 $\pm$ 0.16 & 1.68 $\pm$ 0.21 \\
        53230  &  342.199005  &  -44.518162 & 10.12 $\pm$ 0.54 & -13.15 $\pm$ 0.29 & 41.40 $\pm$ 9.85\\
        \hline
        \hline
     \end{tabular}
     \label{tab:z_9_11}
 \end{table*}
 
 \begin{table*}
     \centering
     \caption{Continued the Table \ref{tab:z_9_11}.}
     \begin{tabular}{lccccc}
     \hline
     \hline
       ID & RA(J2000)& DEC(J2000)& z$_{\rm phot}$& \muv [mag]& $\mu$\\
      \hline
     \hline
        2280   &  342.230560  &  -44.567539 & 10.15 $\pm$ 0.27 & -17.85 $\pm$ 0.08 & 1.28 $\pm$ 0.15 \\
        23154  &  342.234375  &  -44.530270 & 10.15 $\pm$ 0.45 & -17.35 $\pm$ 0.14 & 1.68 $\pm$ 0.21 \\
        54189  &  342.198242  &  -44.519115 & 10.16 $\pm$ 0.56 & -12.89 $\pm$ 2.93 & 81.56 $\pm$ 138.65 \\
        2470   &  342.237244  &  -44.566978 & 10.20 $\pm$ 0.33 & -17.50 $\pm$ 0.10 & 1.26 $\pm$ 0.14 \\
        40398  &  342.155487  &  -44.537029 & 10.21 $\pm$ 0.58 & -16.59 $\pm$ 0.18 & 4.12 $\pm$ 0.59 \\
        983    &  342.250427  &  -44.571609 & 10.25 $\pm$ 0.66 & -17.04 $\pm$ 0.19 & 1.21 $\pm$ 0.13 \\
        55874  &  342.190979  &  -44.542007 & 10.46 $\pm$ 4.01 & -15.72 $\pm$ 1.31 & 3.55 $\pm$ 0.53 \\
        \hline
        73273  &  342.236664  &  -44.564953 & 10.56 $\pm$ 0.61 & -16.61 $\pm$ 0.17 & 1.27 $\pm$ 0.14 \\
        18900  &  342.162781  &  -44.528698 & 10.57 $\pm$ 0.63 & -15.42 $\pm$ 0.19 & 3.96 $\pm$ 0.62 \\
        9638   &  342.257690  &  -44.552753 & 10.61 $\pm$ 0.72 & -16.89 $\pm$ 0.20 & 1.26 $\pm$ 0.14 \\
        11545  &  342.221802  &  -44.549793 & 10.61 $\pm$ 0.50 & -16.75 $\pm$ 0.14 & 1.48 $\pm$ 0.17 \\
        55303  &  342.185425  &  -44.520073 & 10.70 $\pm$ 4.21 & -15.12 $\pm$ 1.47 & 14.82 $\pm$ 3.05 \\
        70775  &  342.198120  &  -44.536846 & 10.70 $\pm$ 3.44 & -15.23 $\pm$ 1.08 & 3.77 $\pm$ 0.59 \\
        4305   &  342.217743  &  -44.562569 & 10.82 $\pm$ 0.60 & -16.66 $\pm$ 0.17 & 1.41 $\pm$ 0.16 \\
        41490  &  342.250214  &  -44.538063 & 10.82 $\pm$ 4.06 & -17.43 $\pm$ 1.19 & 1.39 $\pm$ 0.16 \\
        46385  &  342.198792  &  -44.541153 & 10.89 $\pm$ 0.67 & -15.97 $\pm$ 0.19 & 2.62 $\pm$ 0.36 \\
        51235  &  342.190857  &  -44.515186 & 10.91 $\pm$ 0.74 & -15.52 $\pm$ 0.22 & 4.10 $\pm$ 0.61 \\
        13413  &  342.178345  &  -44.547295 & 10.96 $\pm$ 0.35 & -16.11 $\pm$ 0.10 & 3.60 $\pm$ 0.47 \\
        13593  &  342.182129  &  -44.547195 & 10.99 $\pm$ 0.27 & -17.02 $\pm$ 0.08 & 3.04 $\pm$ 0.41 \\
        10172  &  342.244080  &  -44.551899 & 11.00 $\pm$ 1.48 & -17.01 $\pm$ 0.40 & 1.38 $\pm$ 0.16 \\
        69853  &  342.208374  &  -44.530537 & 11.14 $\pm$ 0.82 & -15.82 $\pm$ 0.23 & 3.41 $\pm$ 0.53 \\
        5517   &  342.237762  &  -44.5601200& 11.30 $\pm$ 0.27 & -17.60 $\pm$ 0.07 & 1.29 $\pm$ 0.14 \\
        56130  &  342.194855  &  -44.542469 & 11.32 $\pm$ 4.85 & -15.09 $\pm$ 1.45 & 2.69 $\pm$ 0.38 \\
        72189  &  342.252869  &  -44.545906 & 11.39 $\pm$ 0.43 & -16.90 $\pm$ 0.11 & 1.32 $\pm$ 0.15 \\
        \hline
        6874   &  342.223267  &  -44.557743 & 11.55 $\pm$ 0.21 & -16.98 $\pm$ 0.05 & 1.38 $\pm$ 0.16 \\
        44018  &  342.204620  &  -44.539558 & 11.63 $\pm$ 4.27 & -18.21 $\pm$ 1.14 & 1.26 $\pm$ 0.14 \\
        68320  &  342.177917  &  -44.518085 & 11.74 $\pm$ 0.58 & -16.69 $\pm$ 0.15 & 3.26 $\pm$ 0.52 \\
        6880   &  342.223328  &  -44.557713 & 11.76 $\pm$ 0.23 & -16.67 $\pm$ 0.06 & 1.38 $\pm$ 0.16 \\
        11945  &  342.263031  &  -44.549343 & 11.77 $\pm$ 4.29 & -17.97 $\pm$ 1.13 & 1.26 $\pm$ 0.14 \\
        1781   &  342.239868  &  -44.569035 & 11.85 $\pm$ 0.18 & -16.92 $\pm$ 0.04 & 1.25 $\pm$ 0.14 \\
        4213   &  342.264160  &  -44.562721 & 11.90 $\pm$ 0.81 & -17.26 $\pm$ 0.20 & 1.21 $\pm$ 0.14 \\
        12189  &  342.265320  &  -44.548908 & 11.99 $\pm$ 0.31 & -17.61 $\pm$ 0.08 & 1.29 $\pm$ 0.14 \\
        10313  &  342.192139  &  -44.551685 & 13.78 $\pm$ 5.40 & -16.79 $\pm$ 1.21 & 1.38 $\pm$ 0.16 \\
        57696  &  342.168823  &  -44.521694 & 14.15 $\pm$ 1.77 & -15.32 $\pm$ 0.40 & 5.42 $\pm$ 0.90 \\
        71717  &  342.248749  &  -44.542152 & 14.81 $\pm$ 1.26 & -17.36 $\pm$ 0.24 & 1.36 $\pm$ 0.16 \\
         \hline
         \hline
     \end{tabular}
     \label{tab:z_11_15}
 \end{table*}




\label{lastpage}
\end{document}